


\magnification\magstep1
\parskip=\medskipamount
\hsize=6 truein
\vsize=8.1 truein
\hoffset=.2 truein
\voffset=0.4truein
\baselineskip=14pt
\tolerance=500


\font\titlefont=cmbx12
 at 10 truept
\font\authorfont=cmcsc10
\font\addressfont=cmsl10 at 10 truept
\font\smallbf=cmbx10 at 10 truept
 4



\outer\def\beginsection#1\par{\vskip0pt plus.2\vsize\penalty-150
\vskip0pt plus-.2\vsize\vskip1.2truecm\vskip\parskip
\message{#1}\leftline{\bf#1}\nobreak\smallskip\noindent}


\outer\def\subsection#1\par{\vskip0pt plus.2\vsize\penalty-80
\vskip0pt plus-.2\vsize\vskip0.8truecm\vskip\parskip
\message{#1}\leftline{\it#1}\nobreak\smallskip\noindent}


\newcount\notenumber

\def\note{\advance\notenumber by 1
\footnote{$^{\the \notenumber}$}}      


\newdimen\itemindent \itemindent=13pt
\def\textindent#1{\parindent=\itemindent\let\par=\resetpar%
\indent\llap{#1\enspace}\ignorespaces}

\let\oldpar=\par
\def\resetpar{\oldpar\parindent=0pt\let\par=\oldpar}

\font\ninerm=cmr9 \font\ninesy=cmsy9
\font\eightrm=cmr8 \font\sixrm=cmr6
\font\eighti=cmmi8 \font\sixi=cmmi6
\font\eightsy=cmsy8 \font\sixsy=cmsy6
\font\eightbf=cmbx8 \font\sixbf=cmbx6
\font\eightit=cmti8
\def\eightpoint{\def\rm{\fam0\eightrm}
  \textfont0=\eightrm \scriptfont0=\sixrm \scriptscriptfont0=\fiverm
  \textfont1=\eighti  \scriptfont1=\sixi  \scriptscriptfont1=\fivei
  \textfont2=\eightsy \scriptfont2=\sixsy \scriptscriptfont2=\fivesy
  \textfont3=\tenex   \scriptfont3=\tenex \scriptscriptfont3=\tenex
  \textfont\itfam=\eightit  \def\it{\fam\itfam\eightit}%
  \textfont\bffam=\eightbf  \scriptfont\bffam=\sixbf
  \scriptscriptfont\bffam=\fivebf  \def\bf{\fam\bffam\eightbf}%
  \normalbaselineskip=9pt
  \setbox\strutbox=\hbox{\vrule height7pt depth2pt width0pt}%
  \let\big=\eightbig \normalbaselines\rm}
\catcode`@=11 %
\def\eightbig#1{{\hbox{$\textfont0=\ninerm\textfont2=\ninesy
  \left#1\vbox to6.5pt{}\right.\n@space$}}}
\def\vfootnote#1{\insert\footins\bgroup\eightpoint
  \interlinepenalty=\interfootnotelinepenalty
  \splittopskip=\ht\strutbox %
  \splitmaxdepth=\dp\strutbox %
  \leftskip=0pt \rightskip=0pt \spaceskip=0pt \xspaceskip=0pt
  \textindent{#1}\footstrut\futurelet\next\fo@t}
\catcode`@=12 %


\def\mapright#1{\smash{
  \mathop{\longrightarrow}\limits^{#1}}}
\def\mapdown#1{\Big\downarrow
  \rlap{$\vcenter{\hbox{$\scriptstyle#1$}}$}}
\def\diag{
  \def\normalbaselines{\baselineskip2pt \lineskip3pt
    \lineskiplimit3pt}
  \matrix}


\def\mapright#1{\smash{\mathop{\longrightarrow}\limits^{#1}}}

\def\mapdown#1{\Big\downarrow
  \rlap{$\vcenter{\hbox{$\scriptstyle#1$}}$}}
\def\diag{
  \def\normalbaselines{\baselineskip2pt \lineskip3pt
    \lineskiplimit3pt}
  \matrix}


\def\b{\bar}
\def\h{\hat}
\def\t{\tilde}
\def\l{\langle}
\def\r{\rangle}
\def\A{{\cal A}}
\def\Z{{\cal Z}}
\def\V{{\cal V}}
\def\O{{\cal O}}
\def\H{{\cal H}}

\def\T{{\cal T}}

\def\Hh{\hat{\cal H}}

\def\E{\hbox{$\cal E$}}
\def\F{\hbox{$\cal F$}}
\def\empty{0\!\!\!/}

\def\emu#1#2{\hbox{${\hat e}^{\mu}_{#1#2}$}}
\def\enu#1#2{\hbox{${\hat e}^{\nu}_{#1#2}$}}
\def\Dmu#1#2{\hbox{$D^{\mu}_{#1#2}$}}
\def\Dnu#1#2{\hbox{$D^{\nu}_{#1#2}$}}
\def\vmu#1#2{\hbox{$v^{\mu}_{#1#2}$}}
\def\shalf{\hbox{${\textstyle{1\over 2}}$}}


\rightline{Freiburg THEP-95/16}
\rightline{quant-ph 9508011}
\bigskip
{\baselineskip=24 truept
\titlefont
\centerline{QUANTUM MECHANICS ON SPACES WITH}
\centerline{FINITE FUNDAMENTAL GROUP}
}

\vskip 1.1 truecm plus .3 truecm minus .2 truecm

\centerline{\authorfont Domenico Giulini\footnote*{
e-mail: giulini@sun2.ruf.uni-freiburg.de}}
\vskip 2 truemm
{\baselineskip=12truept
\addressfont
\centerline{Fakult"t f\"ur Physik,
Universit\"at Freiburg}
\centerline{Hermann-Herder Strasse 3, D-79104 Freiburg, Germany}
}
\vskip 1.5 truecm plus .3 truecm minus .2 truecm

\centerline{\smallbf Abstract}
\vskip 1 truemm
{\baselineskip=12truept
\leftskip=3truepc
\rightskip=3truepc
\parindent=0pt

{\eightpoint
We consider in general terms dynamical systems with finite-dimensional,
non-simply connected configuration-spaces. The fundamental group is
assumed to be finite. We analyze in full detail those ambiguities in the
quantization procedure that arise from the non-simply connectedness
of the classical configuration space. We define the quantum theory on
the universal cover but restrict the algebra of observables $\O$ to the
commutant of the algebra generated by deck-transformations. We apply
standard superselection principles and construct the corresponding
sectors. We emphasize the relevance of all sectors and not
just the abelian ones.
\par}}

\beginsection{Section 1. Introduction}

Quantizing a system whose classical configuration space, $Q$, is not
simply connected is ambiguous over and above other ambiguities which
may already be present in the simply connected case. This paper aims
to fully describe and analyze these ambiguities for the cases of
finite fundamental groups without entering any
discussion on problems in quantization proper. For the rest of the
paper we thus assume a definite and consistent prescription for
quantization on simply connected configuration spaces (or at least
specific examples thereof, e.g. homogeneous spaces) to exist and
focus attention to the {\it additional} ambiguities in the non
simply-connected case. We are interested in non-abelian fundamental
groups and, necessarily, their representation theory. It is to evade
the unfortunate intricacies of representation theory for infinite
discrete non-abelian groups that we restrict attention to finite groups.
This at least allows a general treatment, although there are certainly
many cases where specific infinite groups are of interest.

{}From the technical point of view the ambiguities we are interested in
appear in a variety of guises, depending in particular on the
quantization scheme that is employed. For example, attempting standard
canonical quantization rules on $R^2-\{0\}$ (the famous Bohm-Aharonov
situation) results in unitarily inequivalent representations of the
canonical commutation relations [Re]. This is possible since the point
defect and its associated incompleteness prevent the representations
to exponentiate to the Weyl form of the commutation relations and
therefore the application of von Neumann's well known uniqueness
result ([RS], theorem VIII.14). An even simpler situation
that captures all the essential features involved here is given by
a particle on the circle (compare remark 3.1.6;5 in [T]).

Let us go into some more details by looking at the slightly more
general situation of a particle on the $n$-torus, $T^n$. We
represent the torus by the cube,
$K^n=\{0\leq x_k\leq 1,\, k=1,..,n\}\subset R^n$, whose opposite
sides are eventually identified via translations. For the moment,
however, let us work with the fundamental domain $K^n$. We consider
the Hilbert space $L^2(K^n,d^nx)$ and in it the dense domain of
absolutely continuous functions, $\psi$, which vanish on the
boundary $\partial K^n$, and whose first derivatives are again
in the Hilbert space. The momentum operators,
$p_k=-i{\partial\over \partial x_k}$, are not self-adjoint on
this domain but admit self-adjoint extensions by relaxing the
boundary conditions to
$\psi\vert_{x_k=1}=\exp(i\theta_k)\psi\vert_{x_k=0}$, where
each $\theta_k$ is some absolutely continuous but otherwise
arbitrary function of the $n-1$ variables $x_i$, $i\not =k$.
Each of the now self-adjoint operators $p_k$ (we shall use the
same symbol) exponentiates to a one-parameter unitary group:
$R\ni a\rightarrow \exp(iap_k)=U_k(a)$,
where $U_k(a)$ displaces $\psi$ by an amount $a$ in the positive
$x_k$-direction so that values that are pushed through the boundary
$x_k=1$ reenter at $x_k=0$ with the additional phase
$\exp(-i\theta_k)$. At this point we note that our self-adjoint
extensions are too general, since for non-constant $\theta_k$ the
unitaries $U_k(a)$, and hence the $p_k$, will not mutually
commute (compare section VIII.5 in [RS]).
Since we want our extensions $p_k$ to commute we restrict to
constant $\theta_k$. The inequivalent commuting extensions for
the momenta are thus labelled by $n$ angles $\theta_1,\dots, \theta_n$.
If we finally identify opposite faces of $K^n$ so as to obtain the
$n$-torus, $T^n$, all the inequivalent quantizations still persist if
we allow the `functions' $\psi$ to be sections in flat complex
line-bundles-with-connection over $T^n$ [Wo]. The fundamental
group of $T^n$ is ${\bf Z}^n$, and the flat line-bundles-with-connection
are classified by the inequivalent one-dimensional irreducible
representations thereof (see e.g chapter 5 in [Wo]). These are just
labelled by the angles $\theta_1,\dots,\theta_n$ whose
interpretation in the bundle picture is to fix the representation
for the transition functions and also to determine the holonomies:
$\exp(i\theta_k)$ is the holonomy for the loop along the $x_k$
coordinate.

{}From this example it should be clear that the geometric picture
underlying the possibility of inequivalent quantizations is fairly
simple. It is therefore not surprising that these possibilities
were first systematically studied within the path-integral
formulation [LD], where different homotopy classes of paths
connecting two fixed points need not carry the same weight in
the path integral. (See also [Sch] for an early discussion.)
Rather, they could carry relative weights given by complex numbers
of unit modulus. Unitarity then implies that these weight factors must
furnish some one-dimensional complex unitary representation of the
fundamental group. This prescription is most conveniently formulated
by employing the universal cover, $\b Q$, of the configuration space
$Q$ as domain for the quantum mechanical state function [Do1-2].
At least in the case of finite coverings one may then  simply work
on the universal cover space. The redundancy it represents is
restricted to finitely many repetitions which can easily be
accounted for by appropriate normalization factors. In the case of
infinite groups one may select a fundamental domain $\b F\subset\b Q$
for $Q$ and chose the Hilbert space to be square integrable
functions on $\b F$. This is precisely what we did in the torus
example above. However,
in the sequel we restrict to finite coverings and here $\b Q$ is more
convenient to work with than $\b F$. Any quantum mechanical system
based on $Q$ can be lifted to define such a system on $\b Q$ so that
all the operations may now be carried out on the simply connected
space $\b Q$. The distinguishing feature of a quantum mechanical
system so obtained from a system with genuine classical configuration
space $\b Q$ is the absence of certain observables in the former case.
For example, disjoint sets on $\b Q$ which cover the same set on $Q$
cannot give rise to different projection operators, as it would be
the case if we considered a system whose configuration space were
truly given by $\b Q$. Hence the idea is that due to
missing observables we encounter superselection rules, and that the
quantization ambiguities are precisely given by the different sectors.
We stress that we wish to consider {\it all} sectors arising in this
fashion.

The plan of the paper is as follows: In section 1 we outline the
underlying classical geometry thereby introducing some notation.
In this setting we briefly review the known case where the
fundamental group is abelian [LD].
Section 2 presents in an explicit way the geometry of the regular
representation for general finite groups. In section 3 we use a
finite-dimensional Hilbert space with reducible algebra of observables
as a toy model to introduce some basic concepts from the
theory of superselection rules in ordinary quantum mechanics.
In section 4 we finally generalize the constructions mentioned in
section 1 to the non-abelian case. We show how to implement the
requirement of so-called abelian superselection rules which in the non
abelian case is not automatic. Coherent sectors are built from sections in
vector bundles for each irreducible representation of the fundamental
group. Appendix A provides some explanation on how gauge theoretic
concepts  apply to the universal cover space and its associated vector
bundles. Appendix B contains a simple quantum mechanical example
with non-abelian finite fundamental group. Throughout this paper
we shall not employ the summation convention for repeated indices.

\beginsection{Section 1. Classical Background and Abelian Case}

Let $Q$ be a finite-dimensional manifold that serves as configuration
space for some dynamical system. We denote its cotangent bundle by
$T^*(Q)$. $\pi_1(Q,q)$ denotes the fundamental group of $Q$ based at the
point $q$. It is assumed to be finite, and hence for each $q$ abstractly
isomorphic to a finite group $G$. The neutral element of $G$ will be
called $e$. We stress that although there exist isomorphisms of
$\pi_1(Q,q)$ with $G$ for each $q$, there are generally no natural
choices for these isomorphisms and hence no natural identifications
of the fundamental groups at various points with $G$ (see appendix A).
There are, however, natural identifications of the conjugacy classes
of each $\pi_1(Q,q)$ with those of $G$. Abelian fundamental groups
may thus be identified with an abstract abelian group.
In this case it makes sense to speak of {\it its} (meaning $Q$'s)
fundamental group, a terminology which otherwise just refers to an
abstract isomorphism. The relevance of this point to our discussion
should not be overlooked (compare appendix A).

Let further $\b Q$ denote the universal covering manifold and
$\tau:\b Q\rightarrow Q$ the projection map. Points of $\b Q$
are denoted by $\b q,\b p,$ etc., where sometimes we use this
notation to also indicate that $\tau(\b q)=q$ etc.. $\b Q$ has
the structure of a $G$-principal bundle:
$$
\diag{
G&\mapright{}&\b Q \cr
&&\mapdown{\tau}   \cr
&&Q                \cr}
\eqno{(1.1)}
$$
where $G$ acts on $\b Q$ from the right:
$$\eqalignno{
& G\times \b Q\rightarrow \b Q\,,\quad
(g,\b q)\mapsto R_g(\b q)=:\b qg\,,          &(1.2) \cr
& \hbox{such that}\quad \tau\circ R_g=
\tau\quad\forall g\in G\,.                   &(1.3) \cr}
$$
Since $G$ is discrete, $\tau$ is a local diffeomorphism and the
tangent maps $\tau_{\b q\,*} :\,T_{\b q}(\b Q)\rightarrow T_q(Q)$ are linear
isomorphisms with inverse
$\tau^{-1}_{\b q\,*}:\,T_q(Q)\rightarrow T_{\b q}(\b Q)$ for each
$\b q\in\b Q$. For them (1.3) implies:
$$
\left( R_{g^{-1}}\right)_{\b q g\,*}\circ\tau^{-1}_{\b q g\,*}
=\tau^{-1}_{\b q\,*}\,.
\eqno{(1.4)}
$$
We can now lift $\tau$ to the cotangent bundles $T^*(\b Q)$ and $T^*(Q)$
of $\b Q$ and $Q$ (call the lift $\t \tau$) and combine it with the
natural lift, ${\t R}_g$, of $R_g$ into the following diagram with two
commuting squares:
$$
\diag{
T^*(\b Q)&\mapright{{\t R}_g}&T^*(\b Q)&\mapright{\t \tau}&T^*(Q) \cr
\mapdown{\b \pi}&&\mapdown{\b \pi}&&\mapdown{\pi}                 \cr
\b Q&\mapright{R_g}&\b Q&\mapright{\tau}&Q                           \cr}
\eqno{(1.5)}
$$
We denote points of the cotangent bundle by greek letters with
occasionally added subscripts indicating their base point. We have
$$
\eqalignno{
& \t R_g({\b \alpha}_{\b q}):=
{\b \alpha}_{\b q}\circ \left(R_{g^{-1}}\right)_{\b q g\, *}
\quad \forall {\b \alpha}_{\b q}\in T^*_{\b q}(\b Q)\,,
& (1.6)\cr
& \t \tau({\b \alpha}_{\b q}):={\b \alpha}_{\b q}\circ\tau^{-1}_{\b q\,*}
\quad\forall{\b \alpha}_{\b q}\in T^*_{\b q}(\b Q)\,,
&(1.7)\cr}
$$
so that, using (1.4), we get in analogy to $(1.3)$:
$$
\t \tau\circ\t R_g=\t \tau \,.
\eqno{(1.8)}
$$

Let ${\b \alpha}_{\b q}\in T_{\b q}^*(\b Q)$ and
$\alpha_q\in T_q^*(Q)$, so that ${\t \tau}({\b \alpha})=\alpha$,
i.e., $\alpha_q\circ\tau_{\b q\,*}={\b \alpha}_{\b q}$.
The canonical 1-forms on $T^*(\b Q)$ and $T^*(Q)$ are defined by
${\b \sigma}_{{\b \alpha}}:={\b \alpha}\circ\b \pi_*$ and
$\sigma_{\alpha}:=\alpha\circ\pi_*$ respectively. Then
$$
  {\t \tau}^*_{{\b \alpha}} (\sigma_{\alpha})
= \sigma_{\alpha}\circ{\t \tau}_{\b \alpha\, *}
= \alpha\circ\pi_{\alpha\, *}\circ{\t \tau}_{\b \alpha\, *}
= \alpha\circ\tau_{\b q\, *}\circ{\b \pi}_{\b \alpha\, *}
= {\b \sigma}_{\b \alpha}
\eqno{(1.9)}
$$
so that $\t \tau$ is exact-symplectic. The same holds obviously for
all ${\t R}_g$, so that phase space functions invariant under all
${\t R}_g$ generate an invariant flow on $T^*(\b Q)$. It is easy
to see that ${\t R}_g$-invariant ($\forall g\in G$) functions $\b H$ on
$T^*(\b Q)$ are precisely those of the form $\b H=H\circ{\t \tau}$,
where $H$ is a function on $T^*(Q)$. Given such a function as a
Hamiltonian, the dynamical descriptions using $(T^*(Q),H)$ and
$(T^*(\b Q),\b H)$ are equivalent in the following sense: pick
$\alpha\in T^*(Q)$ and any $\b \alpha\in T^*(\b Q)$ satisfying
${\t \tau}(\b \alpha)=\alpha$. Let $\b \gamma (t)$ be the uniquely
determined solution curve on $T^*(\b Q)$ for the Hamiltonian $\b H$
which satisfies $\b \gamma (t=0)=\b \alpha$. Then
$\t \tau\circ\b \gamma=\gamma$, where $\gamma$ is the unique solution
curve on $T^*(Q)$ for the Hamiltonian $H$, satisfying
$\gamma(t=0)=\alpha$. In this way, the Hamiltonian description on
$T^*(\b Q)$ using only observables of the form
$$
\b O=O\circ \t \tau
\eqno{(1.10)}
$$
is entirely equivalent to the description on $T^*(Q)$.
Note that generally the maps $\tau^{-1}_{\b q\,*}$ allow to
uniquely lift any vector field $X$ on $T^*(Q)$ to a vector
field $\b X$ on $T^*(\b Q)$ which is invariant under the action
${\t R}$ of $G$. (The same holds, of course, for vector fields
on $Q$ and $\b Q$.) Moreover, $\b X$ is locally Hamiltonian if
$X$ is. The converse is not quite true, since it might happen
that for some properly locally Hamiltonian $X$ its lift, $\b X$,
is in fact globally Hamiltonian. It is obvious that $\b X$ is
complete if $X$ is. If a (symmetry-) group $S$ acts on $T^*(Q)$
it will generally not be true that it also acts on $T^*(\b Q)$.
For example, let the vector field $X$ on $T^*(Q)$ generate the
circle group and suppose that its orbit loops are not
contractible\note{For connected $T^*(\b Q)$ either all or none
of the orbits are contractible.}.
Then it is clear that only a cover group of the circle will act on
$T^*(\b Q)$. Generally, there will be an action of a larger group,
$S_G$, given by some $G$-extension of $S$\note{$G$ is a normal
subgroup of $S_G$ so that $S_G/G\cong S$. But if $S$ is not a
subgroup of $S_G$ there will be no action of $S$ on $T^*(\b Q)$.
Since we consider only finite $G$, $S_G$ will be compact if $S$ is.}.

Let us now turn to the quantization, where the Hilbert space is
built from square integrable complex functions on $\b Q$. The
measure $d\b q$ on $\b Q$ is taken as the pullback of the
measure $dq$ on $Q$ via $\tau$, so that, $\forall g\in G$,
$$
R^*_gd\b q=d\b q\,.
\eqno{(1.11)}
$$
In analogy to the classical case, we require: {\it observables must
commute with the action of $G$ on $L^2(\b Q,d\b q)$}. For example,
integral kernels of operators on $L^2(\b Q,d\b q)$ which satisfy
$$
\b O(\b q g,\b p g)=\b O(\b q,\b p)\quad
\forall\b q,\b p\in\b Q, \,\,\forall g\in G
\eqno{(1.12)}
$$
clearly commute with the action of $G$. In particular this is true for
the propagator:
$$
\b K(\b q',t';\b q,t)=\b K(\b q'g,t';\b qg,t)\,.
\eqno{(1.13)}
$$
In [LD][D1-2] it was pointed out that the wave function on $\b Q$
need not project to a well defined function on $Q$. Rather, one
could also consider wave functions that satisfied
$$
\psi^{\mu}(\b q g)=\chi^{\mu}(g)\psi^{\mu}(\b q),
\eqno{(1.14)}
$$
where $\mu$ labels a one-dimensional complex unitary irreducible
representation of $G$ with characters $\chi^{\mu}(g)$. On $Q$ such
wave functions are sections in a complex line bundle which is
$\chi^{\mu}$-associated to the principal bundle (1.1). In general we
prefer however to work instead with functions on $\b Q$ satisfying
(1.14), called the condition of $\chi^{\mu}$-equivariance (compare
appendix A).
We thus have the Hilbert spaces $\H=L^2(\b Q,d\b q)$ and the
subspaces $\H^{\mu}$ of those functions satisfying $(1.14)$.
A key point is now to establish that the observables
act indeed irreducibly on each $\H^{\mu}$. This will follow from
a more general result proven in chapter 4.

Let us consider the operator
$$\eqalign{
T^{\mu}&:\H\rightarrow\H^{\mu}                      \cr
\left(T^{\mu}\psi\right)(\b q)&:=
  {1\over n}\sum_{g\in G}\chi^{\mu}(g)\psi(\b q g),  \cr}
\eqno{(1.15)}
$$
which is easily seen to to be self-adjoint. It satisfies
$$
T^{\mu} T^{\nu}=\delta_{\mu\nu}T^{\mu}
\eqno{(1.16)}
$$
due to the orthogonality of the characters. Moreover, $T^{\mu}$ restricts
to the identity on $\H^{\mu}$. The set $\{T^{\mu}\}$ is thus just the
collection of projection operators onto the mutually orthogonal subspaces
$\{\H^{\mu}\}$ of $\H$. Since the propagator satisfies (1.13), we
have\note{Composition of maps will generally be denoted by the symbol
$\circ$. In the very obvious cases it will be omitted, like in (1.16).}

$$
T^{\mu}\circ \b K(t';t)=\b K(t';t)\circ T^{\mu}=
T^{\mu}\circ\b K(t';t)\circ T^{\mu}=:{\b K}^{\mu}(t',t)\,,
\eqno{(1.17)}
$$
where explicitly
$$
{\b K}^{\mu}(\b q',t';\b q,t)=
{1\over n}\sum_{g\in G}\chi^{\mu}(g)\b K(\b q' g,t';\b q,t)\,.
\eqno{(1.18)}
$$
The standard combination property for propagators, satisfied by
$\b K$, now implies the same for each $\b K^{\mu}$:
$$
\int_{\b Q} \b K^{\mu}(\b q',t';\b q'',t'')\,
\b K^{\mu}(\b q'',t'';\b q,t)\,d\b q'' =
\b K^{\mu}(\b q',t';\b q,t)\,.
\eqno{(1.19)}
$$
Finally, we note that due to (1.12) formulae (1.17-1.18) identically hold
when $\b K$ is replaced with $\b O$:
$$\eqalign{
& T^{\mu}\circ \b O=\b O\circ T^{\mu}=
  T^{\mu}\circ \b O\circ T^{\mu}=:{\b O}^{\mu}       \cr
& {\b O}^{\mu}(\b q,\b p)={1\over n}\sum_{g\in G}
  \chi^{\mu}(g)\b O(\b q g,\b p)                     \cr}
\eqno{(1.20)}
$$

This is essentially the framework of [LD][D1-2]. We believe,
however, that starting from (1.14) (or (1.18)) is a rather ad
hoc procedure and that the actual task is to construct
{\it all} subspaces of $\H$ in which observables act irreducibly.
This is not achieved by considering all $\H^{\mu}$, since generally
$$
\H\not =\bigoplus_{\mu={\rm 1-dim}} \H^{\mu}\,.
\eqno{(1.21)}
$$
Only for abelian groups could the equality sign hold in (1.21).
In section 4 we give the generalization to non-abelian
finite groups $G$. Similar ideas how this could be done were also
formulated in a non-technical fashion in [So] and [Ba1-2].
But before attacking the actual problem,  we need to present some
standard facts about the regular representation of finite groups.
This will be done in some detail in the next section.

\beginsection{Section 2. The Geometry of the Regular Representation}

Let $G$ be a finite group of order $n$ and unit element $e$.
The group algebra $V_G$ is the vector space
$$
V_G:=\hbox{span}\left\{\h g,\, g\in G \right\}
\eqno{(2.1)}
$$
where from now on a hat identifies an element of $V_G$. $V_G$ is
made into an algebra by the obvious multiplication law on the
basis vectors:
$$
\h g\cdot\h h:=\widehat{g\cdot h},
\eqno{(2.2)}
$$
and linear extension. Given any two elements $\h v$ and $\h w$ of
$V_G$,
$$
\h v=\sum_{g\in G} v(g)\h g\,, \quad
\h w=\sum_{g\in G} w(g)\h g \,,\quad v(g),w(g)\in C,
\eqno{(2.3)}
$$
the components of their product are hence given by
$$
(\h v\cdot\h w)(g)=\sum_{h\in G} v(gh^{-1})w(h)=
                   \sum_{h\in G} v(h)w(h^{-1}g).
\eqno{(2.4)}
$$
The algebra $V_G$ is called the group algebra of $G$ and the
representations of $G$ on $V_G$ by left or right multiplication
are called the left or right regular representation respectively.
Under such a regular representation $V_G$ decomposes as
(see [We] for a general discussion)
$$\eqalignno{
&V_G=\bigoplus_{\mu=1}^m V^{\mu}\quad \quad
\hbox{(uniquely),}                                       &(2.5a)\cr
&V^{\mu}=\bigoplus_{i=1}^{n_{\mu}} V^{\mu}_{i,\{L,R\}}\quad
\hbox{(non uniquely),}                                   &(2.5b)\cr}
$$
where $\mu=1,\dots, m$ labels all the inequivalent irreducible
representations of $G$, and $i=1,\dots,n_{\mu}$ labels the
copies of the $\mu$-th representation. $\{L,R\}$ is understood
to replace either $L$ or $R$. $V^{\mu}_{i,\{L,R\}}$ are irreducible
subspaces for the left $(L)$ and right $(R)$ multiplications
respectively. As indicated, for neither of them the decomposition
of $V^{\mu}$ is unique, whereas the decomposition of $V_G$ into the
$V^{\mu}$ is unique. This will become more transparent as we proceed.
It is a property of the regular representation that it contains
each irreducible representation as often as its dimension,
that is, $n_{\mu}=\hbox{dim}V^{\mu}_i$-times (e.g. [We][Ha]).
Hence
$$
\hbox{dim}V^{\mu}=n_{\mu}^2\quad \hbox{and} \quad
\sum_{\mu=1}^m n_{\mu}=n.
\eqno{(2.6)}
$$
Performing left and right $G$-multiplications simultaneously, we obtain
a left $G\times G$-action on $V_G$:
$$
\left((g,h),\h v\right)\mapsto \h g\cdot\h v\cdot \h h^{-1},
\eqno{(2.7a)}
$$
which, by linear extension, yields an action of the corresponding group
algebra $V_{G\times G}\cong V_G\otimes V_G$ on $V_G$:
$$
V_{G\times G}\times V_G \ni
\left(\sum_{g,h}\alpha(g,h)\,\h g\otimes \h h\,,\,\h v\right)
\mathop{\longmapsto}^{\rho}\sum_{g,h}\alpha(g,h)\,\h g\cdot\h v
\cdot \h h^{-1} \in V_G.
\eqno{(2.7b)}
$$
The algebras of left and right multiplications are contained in
$V_{G\times G}$ as subalgebras $V_G\otimes {\h e}$ and
${\h e}\otimes V_G$ respectively, with centralizers
$V^c_G\otimes V_G$ and $V_G\otimes V^c_G$, where $V^c_G$ denotes
the centers of $V_G$. The images of these centralizers under
$\rho$ are isomorphic to $V_G$.

For what follows it will be convenient to employ a special basis
of $V_G$ which is adapted to the decomposition (2.5). We construct
it by assuming we are given a complete set of unitary irreducible
representation matrices $D^{\mu}_{ij}(g)$. Special choices
within the unitary equivalence class can be made if required.
By virtue of the orthogonality relations (e.g. [Ha]),
$$\eqalignno{
{n_{\mu}\over n}\sum_{g}\Dmu ij(g^{-1})D^{\nu}_{kl}(g)
& = \delta_{\mu\nu}\delta_{il}\delta_{jk}, &(2.8a)\cr
\sum_{\mu, i,j}{n_\mu\over n} D^{\mu}_{ij}(g^{-1})D^{\mu}_{ji}(h)
& = \delta_{gh},                            &(2.8b)\cr}
$$
we can use the $\Dmu ij$ as coefficients for a new basis,
$\{\emu ij\}$, of $V_G$, defined by
$$\eqalignno{
\emu ij:=&{n_{\mu}\over n}\sum_g \Dmu ij(g^{-1})\,\h g\,,
 &(2.9a)\cr
\hbox{and inversely}\quad \h g=&\sum_{\mu, i,j}\emu ij \Dmu ji(g)\,.
&(2.9b)\cr}
$$
With respect to these two bases a general element $\h v\in V_G$
has the expansions
$$
\h v=\sum_g v(g)\h g=\sum_{\mu,i,j}v^{\mu}_{ij}\emu ji
\eqno{(2.10a,b)}
$$
and from (2.9) we infer the transformation rules for the components
$$\eqalignno{
v^{\mu}_{ij}&=\sum_g v(g)D^{\mu}_{ij}(g)\,,
 &(2.11a)\cr
v(g)&=\sum_{\mu,i,j}{n_{\mu}\over n}v^{\mu}_{ij}D^{\mu}_{ji}(g^{-1})\,.
 &(2.11b)\cr}
$$
Left and right $\h h$-multiplications are now given by
$$\eqalignno{
\h h\cdot\emu ij &= \sum_k \emu ik \Dmu kj(h) ,  &(2.12a)\cr
\emu ij\cdot\h h &= \sum_k \Dmu ik(h) \emu kj .  &(2.12b)\cr}
$$
The rows and columns of $\emu ij$, considered as a matrix in $ij$,
thus span left- and right-irreducible subspaces respectively, which
we may take as our $V^{\mu}_{i,L}$ and $V^{\mu}_{i,R}$ in the decomposition
$(2.5b)$. For the algebra $V_G$ this means that
$$\eqalignno{
V^{\mu}_{i,L} &= \hbox{span}\,\{\emu i1,\dots,\emu i{n_{\mu}}\}\quad
\hbox{is a minimal left ideal}, &(2.13a)\cr
V^{\mu}_{i,R} &= \hbox{span}\,\{\emu 1i,\dots,\emu {n_{\mu}}i\}\quad
\hbox{is a minimal right ideal}, &(2.13b)\cr
V^{\mu} &=\bigoplus_i^{n_{\mu}} V^{\mu}_{i,\{L,R\}}\quad
\hbox{is a minimal 2-sided ideal}. &(2.13c)\cr}
$$
In terms of the basis $\{\emu ij\}$ the multiplication law can be
easily inferred from $(2.8a)$ and $(2.9a)$:
$$
\emu ij\cdot\enu kl = \delta_{\mu\nu}\delta_{il}\emu kj ,
\eqno{(2.14)}
$$
which implies that components (compare $(2.10b)$) just multiply like
matrices:
$$
\left(\h v\cdot\h w\right)^{\mu}_{ij}=
\sum_k v^{\mu}_{ik}w^{\mu}_{kj} .
\eqno{(2.15)}
$$
Left and right multiplications by $\emu ik$ are then given by
$$\eqalignno{
\emu ik\cdot\h v &=\sum_j v^{\mu}_{ij}\emu jk , &(2.16a)\cr
\h v\cdot\emu ik &=\sum_j v^{\mu}_{jk}\emu ij , &(2.16b)\cr}
$$
which, in an obvious sense, say that left/right multiplication by $\emu ik$
results in writing the content of $V^{\mu}_{i,R}/V^{\mu}_{k,L}$ into
$V^{\mu}_{k,R}/V^{\mu}_{i,L}$ and deletion of all other components.

Let us define ${\h e}^{\mu}_i:=\emu ii$ and
${\h e}^{\mu} :=\sum_i\emu ii$. It follows from $(2.9b)$ that
$\h e=\sum_{\mu}{\h e}^{\mu}$. The spaces $V^{\mu}$ form subalgebras with
units ${\h e}^{\mu}$. Left/right multiplication by ${\h e}^{\mu}_j$
correspond to projection into $V^{\mu}_{j,R}/V^{\mu}_{j,L}$, as is easily
seen from the following special cases of $(2.14)$ and
$(2.9b)$:
$$\eqalignno{
{\h e}^{\mu}_j\cdot\enu ik
 &= \delta_{\mu\nu}\delta_{kj}\enu ik ,         &(2.17a)\cr
\enu ik\cdot{\h e}^{\mu}_j
 &= \delta_{\mu\nu}\delta_{ij}\enu ik ,         &(2.17b)\cr
{\h e}^{\mu}_i\cdot{\h e}^{\nu}_j
 &= \delta_{\mu\nu}\delta_{ij}{\h e}^{\mu}_i ,  &(2.18)\cr
\sum_{\mu, i}{\h e}^{\mu}_i
 &= \h e .                                      &(2.19)\cr}
$$
The projection into $V^{\mu}$ is given by right or left
multiplication with ${\h e}^{\mu}$. It follows that
$$
A:=\hbox{span}\{{\h e}^1_1,\dots, {\h e}^1_{n_1},\dots\dots,
                 {\h e}^m_1,\dots,{\h e}^m_{n_m}\}
\eqno{(2.20)}
$$
is a maximal abelian subalgebra of $V_G$ of dimension
$\sum_{\mu=1}^m n_{\mu}$. Indeed, commutativity of $\h v\in V_G$ with
all elements of $A$ implies that its projection into $V^{\mu}_{i,R}$
equals its projection into $V^{\mu}_{i,L}$ for all $i$. But the
intersection $V^{\mu}_{i,L}\cap V^{\mu}_{i,R}$ is the ray spanned by
${\h e}^{\mu}_i$. Thus $\h v$ must be in $A$ which shows maximality.
In comparison, the centre $V^c_G$ of $V_G$ is also easily determined,
for $\h v\cdot\h g=\h g\cdot\h v$ $\forall g\in G$ implies via $(2.12)$
that $\sum_i \vmu ji\Dmu ik(g)=\sum_i\Dmu ji(g)\vmu ik$,
$\forall g\in G$. Schur's Lemma then yields $\vmu ij=v^{\mu}\delta_{ij}$,
so that
$$
V^c_G=\hbox{span}\{{\h e}^1, \dots, {\h e}^m\}.
\eqno{(2.21)}
$$
Note that, unless $G$ is abelian, the centre of the group algebra
contains but is not equal to the group algebra of the centre, $G_c$,
of $G$. For example, for non-abelian $G$, $\sum_g\h g$ is in $V_G^c$
but not in the group algebra of $G_c$.

The projection maps
$\h v\mapsto {\h e}^{\mu}\cdot \h v=\h v\cdot {\h e}^{\mu}$ are
homomorphisms from $V_G$ onto the subalgebras $V^{\mu}$. Left and right
actions of $V_G$ on the $V^{\mu}$'s thus factor through these projections.
The centralizers $Z^{\mu}$ of $V^{\mu}$ are easily seen to
be given by the subalgebras
$$
Z^{\mu}=\left\{\bigoplus_{\nu\not =\mu}V^{\nu}\right\}
               \oplus\hbox{span}\left\{{\h e}^{\mu}\right\}\,.
\eqno{(2.22)}
$$
Obviously we have $Z^{\mu}=V_G$, iff the $\mu$-th representation is
abelian. From $(2.9)$ it follows that $\h g\in Z^{\mu}$, iff
$D^{\mu}_{ij}(g)=\eta(g)\delta_{ij}$. This is the case iff
$g\in C^{\mu}:=\{h\in G\,/\,D^{\mu}(hf)=D^{\mu}(fh)\ \forall f\in G\}$.
$C^{\mu}$ is a normal subgroup of $G$ and
$D^{\mu}(g)=D^{\mu}(g_G)$ for any $g\in C^{\mu}$, where
$g_G\subset C^{\mu}$ denotes the conjugacy class of $g$ in $G$.

Whereas $V_G$ decomposes unambiguously into the $V^{\mu}$'s for both
left and right multiplication, our choice of left and right
irreducible subspaces $V^{\mu}_{i,\{R,L\}}$ is not unique.
To see what freedom there is, we prove the following

\proclaim Lemma. Let $\h v\in V_G$. The following statements are
                 equivalent. (i) $\h v$ lies in a left-irreducible
                 subspace, (ii) $\h v$ lies in a right irreducible
                 subspace, (iii) $\h v$ has expansion coefficients
                 $\vmu ij=\delta_{\mu\nu}a_ib_j$ for some complex
                 valued $n_{\mu}$-tuples $\{a_i\}$ and $\{b_i\}$.
                 \par

\noindent{\bf Proof.} We shall only prove $(i)\Leftrightarrow (iii)$
since $(ii)\Leftrightarrow (iii)$ is entirely analogous.
$(iii)\Rightarrow (i)$ is trivial. Conversely, assuming that $\h v$
lies in a left-irreducible subspace, we know from $(2.5b)$ that it
must lie in an $n_{\nu}$-dimensional subspace of some $V^{\nu}$, which
for the moment we call $L$. This explains the $\delta_{\mu\nu}$ in
$(iii)$. We set $\h v=\sum_{i,j}v_{ij}\enu ji$. Left multiplications
by $\enu kl$ for all $k,l\in\{1,\dots,n_{\nu}\}$ produces the
$n_{\nu}\times\hbox{rank}\{v_{ij}\}$ linearly independent vectors
$\enu kl\cdot\h v=\sum_j v_{kj}\enu jl$ in $L$. But $L$ is only
$n_{\nu}$-dimensional so that $\hbox{rank}\{v_{ij}\}=1\Leftrightarrow
v_{ij}=a_ib_j\,\bullet$

This shows that any other adapted basis, i.e., where each basis vector
lies in an irreducible subspace, is necessarily of the form (matrix
notation)
$$
{\h \eta}^{\mu}=M{\h e}^{\mu}N^{-1}\,,\quad M,N\in GL(n_{\mu},C),
\eqno{(2.23)}
$$
so that the left and right actions of $G$ are now represented
equivalently to $(2.12)$:
$$\eqalignno{
\h g\cdot{\h \eta}^{\mu} &= {\h \eta}^{\mu}(ND^{\mu}(g)N^{-1}),
                         &  (2.24a)\cr
{\h \eta}^{\mu}\cdot\h g &= (MD^{\mu}(g)M^{-1}){\h \eta}^{\mu}.
                         &  (2.24b)\cr}
$$
So far we can therefore stick to any particular choice
of representation matrices in $(2.9a)$.

If we denote by $\{e_i\}$ the standard basis in $C^{n_{\mu}}$, we
can employ the isomorphism
$\sigma:V^{\mu}\rightarrow C^{n_{\mu}}\otimes C^{n_{\mu}}$,
defined by
$$
\sigma(\emu ij):=e_j\otimes e_i ,
\eqno{(2.25)}
$$
to identify $V^{\mu}$ and $C^{n_{\mu}}\otimes C^{n_{\mu}}$ for each $\mu$.
We shall occasionally use this identification without explicitly mentioning
$\sigma$.
As pointed out in $(2.15)$, left and right multiplications then act only
on the left and right $C^{n_{\mu}}$ respectively. From the previous Lemma
we infer that $\h v$ is an element in an irreducible subspace, iff it is
a pure tensor product $a\otimes b$, $a,b\in C^{n_{\mu}}$ for some $\mu$.
This set of pure tensor products (also called rank=1 vectors) is not a
linear space, but contains the linear spaces
$$\eqalignno{
R^{\mu}(a) &:=\hbox{span}\{a\otimes e_1,\dots, a\otimes e_{n_{\mu}}\},
           &  (2.26a)\cr
L^{\mu}(a) &:=\hbox{span}\{e_1\otimes a,\dots, e_{n_{\mu}}\otimes a\},
           &  (2.26b)\cr}
$$
which comprise all the left- and right-irreducible subspaces if $a$ runs
through all of $C^{n_{\mu}}$ and $\mu$ through all values of $1$ to $m$.
Two different vectors $a$ and $a'$ characterize the same irreducible
subspace, iff $a=\alpha a'$ for some $\alpha\in C-\{0\}$. The space of
left- or right-irreducible subspaces within $V^{\mu}$ can thus be
identified with the complex projective space $CP^{n_{\mu}-1}$ of real
dimension $2(n_{\mu}-1)$.

Next we wish to introduce an inner product on $V_G$, denoted by
$\l\cdot\vert\cdot\r$ (antilinear in the first entry). Since right
$V_G$-multiplications will eventually play the r\^ole of gauge
symmetries in our application, we require it to be right invariant.
This leads to the following string of equations (generally an
overbar over $C$-valued quantities denotes
complex-conjugation):
$$\eqalignno{
\l\emu ik\vert\enu lm\r
=& \l\emu ik\cdot g\vert\enu lm\cdot g\r                        &(2.27a)\cr
=& {1\over n}\sum_{g\in G}\l\emu ik\cdot g\vert\enu lm\cdot g\r &(2.27b)\cr
=& {1\over n}\sum_{g\in G}\sum_{r,s}{\b D}^{\mu}_{ir}(g)
   \Dnu ls(g)\l\emu rk\vert\enu sm\r                            &(2.27c)\cr
=& \delta_{\mu\nu}\delta_{il}{1\over n_{\mu}}\sum_r
   \l\emu rk\vert\emu rm\r                                      &(2.27d)\cr
=&:\delta_{\mu\nu}\delta_{il}S^{\mu}_{km},                      &(2.27e)\cr}
$$
where we have used unitarity of the representation matrices $D^{\mu}$
in the second to last step for the first time. So far no choice within
the equivalence class of unitary representations matrices was specified.
A redefinition within the unitary equivalence class implies (matrix
notation)
$$\eqalignno{
D^{\mu} &\mapsto U^{\mu}D^{\mu}(U^{\mu})^{\dagger},           &(2.28a)\cr
{\h e}^{\mu} &\mapsto U^{\mu}{\h e}^{\mu}(U^{\mu})^{\dagger}, &(2.28b)\cr
S^{\mu} &\mapsto U^{\mu}S^{\mu}(U^{\mu})^{\dagger}.           &(2.28c)\cr}
$$
In general we could use it to diagonalize the Hermitean matrix
$S^{\mu}$. We call its eigenvalues $\lambda^{\mu}_k$,
$k=1,\dots, n_{\mu}$, and get from $(2.27)$
$$
\l\emu ik\vert\enu lm\r=
\delta_{\mu\nu}\delta_{il}\delta_{km}\lambda^{\mu}_k\,.
\eqno{(2.29)}
$$
This formula is still completely general. Choosing an inner product
now corresponds to picking $\sum_{\mu=1}^mn_{\mu}$ coefficients
$\lambda^{\mu}_k$. For our later applications we make the
particular choice:
$$
\lambda^{\mu}_k=\lambda^{\mu}={n_{\mu}\over n^2}\,.
\eqno{(2.30)}
$$
Independence of the lower index is in fact a necessary and
sufficient condition to make the right-invariant inner product also
left-invariant. It also means that we actually did not restrict our
choice of unitary representation matrices at all, so that all
redefinitions $(2.28)$ are still at our disposal.
Proportionality of $\lambda^{\mu}$ to $n_{\mu}$
implies that $\h g$ and $\h h$ are orthogonal for for $g\not =h$.
Indeed, using $(2.8b)$ and $(2.9b)$, we obtain
$$
\l\h g\vert\h h\r=\sum_{\mu, j} \Dmu jj(hg^{-1})\lambda^{\mu}=
                 ={1\over n}\delta_{gh}\,.
\eqno{(2.31)}
$$

A linear operator on $V_G$ is said to be right-invariant if its
matrix elements satisfy the analogous condition to $(2.27a)$.
If $O$ is such an operator, we have in analogy to $(2.27)$
$$\eqalignno{
O^{\mu,\nu}_{ik,lm}&:=\l\emu ik\vert O\vert\enu lm\r
=\delta_{\mu\nu}\delta_{il}O^{\mu}_{km},
&(2.32a)\cr
O^{\mu}_{km}&:={1\over n_{\mu}}\sum_r\l\emu rk\vert O\vert\emu rm\r .
&(2.32b)\cr}
$$
On the other hand, using the completeness relation (where we now
employ Dirac's notation of $\vert\hbox{bra}\r$ and $\l\hbox{ket}\vert$
vectors)
$$
{\bf 1}=\sum_{\mu,i,k}\vert\emu ik\r{1\over\lambda^{\mu}}\l\emu ik\vert,
\eqno{(2.33)}
$$
we can write
$$
O={\bf 1}O{\bf 1}=\sum_{\mu,i,k,m}\vert\emu ik\r{1\over\lambda^{\mu}}
\,O^{\mu}_{km}\,{1\over\lambda^{\mu}}\l\emu im\vert,
\eqno{(2.34)}
$$
so that $O$'s action on $\h v$ can be reformulated, using $(2.10b)$
and $(2.29)$, as a left multiplication
$$\eqalignno{
O\h v &= \sum_{\mu,i,k,m}{1\over\lambda^{\mu}}O^{\mu}_{km}\vmu mi\emu ik
             &(2.35a)      \cr
&=\left(\sum_{\mu,i,k}{1\over\lambda^{\mu}}O^{\mu}_{ik}\emu ki\right)
\cdot\h v =:\h o\cdot\h v\,.&(2.35b)\cr}
$$
Clearly, $O$ is Hermitean, iff $O^{\mu}_{ik}={\b O}^{\mu}_{ki}$. $(2.35)$
says that any right-invariant Hermitean operator is given by left
multiplication with an element $\h o\in V_G$ whose coefficients with
respect to the bases $\{\emu ij\}$ and $\{\h g\}$ satisfy respectively
$$
o^{\mu}_{ij}={\b o}^{\mu}_{ji}\,\Leftrightarrow\,
                            o(g)={\b o}(g^{-1})\,.
\eqno{(2.36)}
$$
Since the algebra $V_G$ acts as operators on its underlying vector space,
these last relations have intrinsic meaning on $V_G$ once an inner
product is introduced. In fact, any inner product on $V_G$ defines a
$*$-operation $V_G\rightarrow V_G$, which is antilinear and satisfies
$*\circ *=1$, through, say, left multiplication:
$$
\l\h v\vert {\h o}\cdot \h w\r=:\l{{\h o}^*}\cdot\h v\vert\h w\r \,.
\eqno{(2.37)}
$$
Alternatively, we could have defined the $*$-operation via right
multiplication which in the general case would have led to a different
$*$-map. However, if the inner product is right- and left-invariant,
the two definitions for the $*$-operations agree. In this case it follows
immediately from $(2.37)$ that
$$
{\Big\{}\sum_{g} o(g)\h g{\Big\}}^*=
\sum_{g}{\b o}(g^{-1})\h g \quad\hbox{and}\quad
{\Big\{}\sum_{\mu,i,j}o^{\mu}_{ij}\emu ji{\Big\}}^*=
\sum_{\mu,i,j}{\b o}^{\mu}_{ji}\emu ji\,.
\eqno{(2.38)}
$$
In particular, ${\h g}^*={\h g}^{-1}$ and ${\emu ij}^*=\emu ji$, which,
by $(2.13)$, implies that $\{V^{\mu}\}^*=V^{\mu}$ and
$\{V^{\mu}_{i,R}\}^*=V^{\mu}_{i,L}$.

Algebras with such a $*$-operation are called $H^*$ algebras and elements
invariant under $*$ are called self-adjoint, or Hermitean. The elements
${\h e}^{\mu}_i$ introduced earlier correspond to mutually orthogonal
Hermitean idempotents, as do the elements ${\h e}^{\mu}$. The latter
ones are however decomposable into the former, which are themselves
indecomposable (i.e. so-called primitive idempotents). The subalgebras
$V^{\mu}$ are mutually orthogonal $H^*$ subalgebras. In particular, the
$V^{\mu}$'s are also minimal 2-sided $H^*$ ideals. (The split $(2.5a)$
is thus still valid in the sense of $H^*$ algebras.) In contrast,
since the subspaces $V^{\mu}_{i,\{L,R\}}$ are not invariant under $*$,
they do not form any $H^*$ ideals.

This basically concludes our presentation of the group algebra. In the
fourth section we shall discuss the decomposition of the quantum mechanical
state space according to an inherited $V_G$-action. Not surprisingly,
it will be very similar to the decomposition of $V_G$ under the
regular representation. In fact, we can immediately
build a finite-dimensional toy model with all the essential features.
This we will do first in order to introduce some general concepts and
notations in a simple context. Then we turn to the general quantum
mechanical case.

\beginsection{Section 3. General Concepts and a Toy Model}

Consider the $n$-dimensional Hilbert space $\H=V_G$ with inner product
$\l\cdot\vert\cdot\r$ and the right regular representation of
$V_G$ on $\H$. As we have seen, it is useful with respect to $V_G$'s
action to represent $\H$ as
$$
\H=\bigoplus_{\mu=1}^m\H^{\mu}=
   \bigoplus_{\mu=1}^m C^{n_{\mu}}\otimes C^{n_{\mu}}.
\eqno{(3.1)}
$$
We call $B(\H)$ the algebra of bounded (a redundant adjective in finite
dimensions) linear operators, which here is isomorphic to
the matrix algebra $M(n,C)$. We wish to regard $G$ as a gauge group
with gauge algebra $V_G$, that is, we require observables to commute
with the action of the  group $G$ (such transformations are called
supersymmetries in [JM]). The algebra of observables, $\O$, is thus
defined as the {\it commutant} of (the right-) $V_G$ {\it in}
$B(\H)$, denoted by $V_G'$. Quite generally, given any set
$S\subset B(\H)$, the set of operators commuting elementwise with $S$
forms an algebra, called the commutant, $S'$, of $S$. The double
commutant, $S''$, is easily seen to be just the algebra generated
by $S$. It is stated in $(2.35b)$ that $\O$ is isomorphic to the algebra
of left  $V_G$ multiplications, which, as e.g. expressed by $(2.15)$,
one may identify with a direct sum of matrix algebras:
$$
\O=\bigoplus_{\mu=1}^m M(n_{\mu},C) \,,
\eqno{(3.2)}
$$
where each matrix algebra $M(n_{\mu},C)$ acts on the left
$C^{n_{\mu}}$-factor in $(3.1)$. The representation of $\O$ in
$\H$ is thus highly reducible.
Whenever the algebra of observables is represented in a
reducible fashion, the pair $(\H,\O)$ is said to contain {\it superselection
rules}. In what follows, we shall investigate more into the structure
of these rules. More precisely, we are interested in the geometric
structure of those subsets of $\H$ that represent pure states, where
this has always to be understood {\it relative to} $\O$. As a word of
principle, and as indicated by the word `relative' , we do not wish to
regard states as being attributed with any more status over and
above that which suffices to answer all the questions contained in $\O$.

The centre of $\O$ is the $m$-dimensional algebra generated by the
projection operators, $T^{\mu}:\H\rightarrow \H^{\mu}$, given
by left or right multiplications with ${\h e}^{\mu}$. Obviously,
vectors representing pure states must always lie in some $\H^{\mu}$,
for, given the sum of two nonzero vectors $v\in \H^{\mu}$ and
$w\in\H^{\nu}$, where $\mu\not =\nu$, the density matrix for the pure state
$\vert v\r+\vert w\r$, considered as a positive linear functional
(the expectation value) on $\O$, is identical to the mixed state
$\vert v\r\l v\vert +\vert w\r\l w\vert$. However, the converse is
not true unless $\mu$ labels an abelian representation. Let us therefore
focus on a higher-dimensional $\H^{\mu}$. It may be considered as the
composite state space of two systems, called left and right, with
individual state spaces $C^{n_{\mu}}$, and where we have no observables
for the right system. To say that a vector in $\H^{\mu}$ represent a pure
state now means the following: express it as a density matrix, form the
reduced density matrix for the left system by tracing out the right system,
then this reduced density matrix is pure. We know from elementary quantum
mechanics that this is the case iff the original vector in $\H^{\mu}$ was
a pure tensor product (i.e. of rank one). Taken together with the lemma
above, we arrive at the following statement: a vector in $\H$ represents
a pure state, iff it lies in a left invariant subspace $L^{\mu}(b)$.
We can represent it by a matrix with
components $a_ib_j$. Observables act on the left index, gauge
transformations on the right. $\O$ acts irreducibly on $L^{\mu}(b)$
in which any two rays can be separated by $\O$. However, for each such ray
there is a unique ray in  each $L^{\mu}(b')$, $b'\not =b$, which gives
the {\it same} state for $\O$. We have thus seen that, with respect to $\O$,
the different left invariant subspaces are indistinguishable so that a
pure state is represented by a ray in each left invariant subspace. This is
equivalent to saying that a pure state corresponds uniquely to a whole
right invariant subspace $R^{\mu}(a)$. That higher than one-dimensional
subspaces should represent quantum mechanical states has already been
discussed in the mid 60's in the context of parastatistics
[MG], where these subspaces were called
{\it generalized rays}. There is nothing inconsistent with this
kind of higher-dimensional redundancy. For example, the
superposition principle takes the following
form: three states (generalized rays) $R^{\mu}(a)$, $R^{\mu}(a')$, and
$R^{\mu}(a'')$ are said to be linearly dependent, iff $a$ lies in
the plane determined by $a'$ and $a''$. Alternatively,
given three rays in each left invariant subspace, then the three states
they define are said to be linearly dependent, iff in each left invariant
subspace the rays lie in a plane. It is clear that this is either
simultaneously true in all or none of the subspaces. This definition
coincides with the more abstract prescription given in [Ho].

Although there is nothing wrong with generalized rays, they do seem
to carry unnecessary redundancy as far as the representation of $\O$ is
concerned\note{However, note that $\H$ and $\O$ were not
independently given: $\O$ was defined as the commutant of $V_G$
{\it in} $B(\H)$.}.
This can be expressed in rational terms in a variety of ways.
For example, in ordinary quantum mechanics, one often hears
{\it Dirac's requirement: There exists a complete set of commuting
observables} [Di]. Let us call them $\{A_i\}$.
Here, by definition, completeness means that a set of
simultaneous eigenvalues determine a ray uniquely.
This statement works for finite-dimensional Hilbert spaces but has
to be replaced in infinite dimensions, where, because of
continuous spectra, the proper notion of eigenvectors does not exist.
But this can be cured by a slight reformulation [J]: Let
$\A=\{A_i\}''\subset \O$ be the abelian algebra generated by the set
$\{A_i\}$. The set is said to be complete, iff $\A$ is a maximal
abelian subalgebra\note{The commutant $\{A_i\}'$ is always
a von Neumann algebra, that is, equal to its double commutant.}
of $B(\H)$, that is, iff ${\A}'=\A$. See [J][JM] for more details
and [Wi2] for a recent review. The generally valid replacement for
Dirac's formulation is {\it Jauch's requirement: $\O$ contains a
maximal abelian subalgebra of $B(\H)$}\note{Standard formulations in
the literature usually do not
make this explicit reference to $B(\H)$. We put it to emphasize
the dependence of this statement on $\H$}. It is clear that in our case
the failure to meet these requirements has to do with the existence of
different rays that cannot be separated by $\O$, or equivalently, that $\O$
does not contain all the projectors onto rays representing pure states.
That this is entirely due to the non-commutativity of the gauge group $G$
is made manifest by an equivalent formulation of Jauch's requirement, due
to Wightman [Wi1]. It is also known as the requirement (or hypothesis)
of {\it commutative (or abelian) superselection rules}. We call it
{\it Wightman's requirement: The commutant $\O'$ of $\O$ in $B(\H)$ is
abelian}. We emphasize that $\O$ was assumed to be a von Neumann
algebra\note{The von Neumann property of $\O$ is not necessary to
prove the implication Wightman $\Rightarrow$ Jauch, but for the
converse, therefore showing that without the von Neumann property
Wightman's requirement is logically weaker.}. See e.g. [GMN] for a
simple proof of the equivalence. It tells us that we cannot keep a
non-commutative gauge group if we want to get rid of generalized rays.

Although generalized rays do no harm, they are also not necessary
for the formulation of a quantum mechanical state space incorporating
all the pure states for $\O$. We demonstrate this ``elimination of the
generalized ray'' [HT] in our model, which highlights in an
elementary fashion the last remark of the previous paragraph. The method
is simple: we truncate $\H$ by selecting an $a\in C^{n_{\mu}}$,
say $a=e_1$, and keep only $L^{\mu}(a)=:\H^{\mu}_{\rm tr}$ for each $\mu$.
Within this space we would then have the standard bijection between pure
states and rays representing them. This amounts to truncating the Hilbert
space representing states for $\O$ to
$$
{\H}_{\rm tr}=\bigoplus_{\mu=1}^m {\H}^{\mu}_{\rm tr},
\eqno{(3.3)}
$$
where of course $\H^{\mu}_{\rm tr}=\H^{\mu}$, iff ${\mu}$ is abelian.
Note that no pure state has been lost. Only redundancies have been
eliminated. Pure states are in bijective correspondence with
rays in the subset
$$
\bigcup_{\mu=1}^m \H^{\mu}_{\rm tr}\subset {\H}.
\eqno{(3.4)}
$$
In fact, the space of rays in this subset is just the disjoint union
of the spaces of rays in each $\H^{\mu}_{\rm tr}$.
The representations of $\O$ on $\H$ and $\H_{\rm tr}$ differ
only by trivial multiplicities. In both cases $\O$ is isomorphic to
$$
\O=\bigoplus_{\mu=1}^m{M(n_{\mu}, C)}=
  \bigoplus_{\mu=1}^m B(\H^{\mu}_{\rm tr})\,.
\eqno{(3.5)}
$$
But in the first case each $M(n_{\mu},C)$ appears with multiplicity
$n_{\mu}$. Representations related in this fashion are therefore called
{\it phenomenologically equivalent} [BLOT].
The price for this elimination is that the symmetry group does
not act on ${\H}_{\rm tr}$ anymore. What remains from the gauge algebra
$V_G$ is a residual action of its centre $V_G^c$ which is now generated
by the projections $T^{\mu}: \H_{\rm tr}\rightarrow {\H}^{\mu}_{\rm tr}$.
Clearly the commutant $\O'$ of $\O$ in $B(\H_{\rm tr})$ just satisfies
Wightman's requirement. Equivalently, Jauch's requirement is satisfied,
since projectors onto rays are now all in $\O$ and any abelian
subalgebra generated by a complete set of orthogonal projectors is
maximal in $B(\H_{\rm tr})$. In a sense, $\H$ was too big for $\O$ and
$\H_{\rm tr}$ is the most economical way to represent the pure states
of $\O$. As we have seen, the projectors onto different $L^{\mu}(a)$
were not in $\O$, only the sum of projectors onto the mutually
orthogonal $L^{\mu}(e_i)$ was.

Finally we note that there is a way to satisfy the
Jauch-Wightman requirement {\it and} have the full gauge group $G$
being reduced by the state space, and that is to just truncate the sum
in $(3.3)$ to include only abelian representations.
This in fact is an often adopted point of view since it conforms with
two seemingly obvious requirements. It has e.g. been used to ``prove''
the impossibility of parastatistics in a quantum mechanical framework
[GMN].
In this work we reject this rather ad hoc procedure on the grounds
that it unnecessarily discards the potentially interesting non-abelian
sectors (i.e. those for which $\mu$ labels a non-abelian representation).
For example, non-abelian sectors are in fact used in the theory of
deformed nuclei. This is explained in appendix B. Generally speaking,
it is a perfectly legitimate procedure to use the gauge group to find all
the sectors and then, in order to conform with the Jauch-Wightman
requirement, sacrifice its action up to an abelian residue.
Whoever wants to have the gauge group still acting might work with
generalized rays. This viewpoint is also expressed in [MG] and [HT].

Note that whereas it is true that only the centre of the gauge algebra
acts on $\H_{\rm tr}$ a larger part of it does act on a specific
$\H^{\mu}_{\rm tr}$ considered in isolation. Precisely that subalgebra
of $V_G$ acts on $\H^{\mu}_{\rm tr}$ which commutes with $V_G$ under the
$\mu$-th representation. In the previous section this subalgebra has
been called $Z^{\mu}$ (compare $(2.22)$). As discussed there, the
corresponding part of the gauge group that still acts on
$\H^{\mu}_{\rm tr}$ is given by $C^{\mu}$. The way it acts is obvious,
since commutativity allows us to write it as left-multiplication.

\beginsection{Section 4. The Non-Abelian Case}

As in section 1, we denote by $\H$ the Hilbert space $L^2(\b Q,d\b q)$
with right invariant measure $d\b q$. The right action of $G$ on $\b Q$
induces a right action of $G$ on $\H$, defined by
$$
(g,\psi)\rightarrow T_g\psi:=\psi\circ R_{g^{-1}}.
\eqno{(4.1)}
$$
It is an isometry due to the right-invariance of the measure.
Linear extension yields a right $V_G$-action on $\H$:
$$
(\h v,\psi)\rightarrow T_{\h v}\psi:
=\sum_{g\in G}v(g)\,\psi\circ R_{g^{-1}}.
\eqno{(4.2)}
$$

We also introduce a second Hilbert space, $\Hh$, as completion of
$V_G$-valued, equivariant functions on $\b Q$ which are square
integrable. The point of doing this is that this Hilbert space is
unitarily isomorphic to $\H$ (see (4.9) below) but displays the
representation properties under the action of $V_G$ in a more
direct way. Equivariance means
$$
\h\psi\circ R_g={\h g}^{-1}\cdot\h\psi\,.
\eqno{(4.3)}
$$
The inner product on $\Hh$, denoted by $(\cdot\vert\cdot)$,
is given by
$$
(\h\psi\vert\h\phi):=\int_{\b Q}\l\h\psi(\b q)\vert\h\phi(\b q)\r\,d\b q,
\eqno{(4.4)}
$$
where $\l\cdot\vert\cdot\r$ is defined by $(2.31)$.
Expanding $\h\psi\in\Hh$ in components,
$$
\h\psi=\sum_{h\in G}\h h\,\psi_h,
\eqno{(4.5)}
$$
then (4.3) implies for the component functions
$$
\psi_h\circ R_g=\psi_{gh} \,.
\eqno{(4.6)}
$$

We now define the linear maps
$$\eqalignno{
\F:\, \H & \rightarrow\Hh \,,\quad
    \psi   \mapsto \F(\psi):=\sum_{g\in G}\h g\,\psi\circ R_g, &(4.7a)\cr
\E:\,\Hh & \rightarrow\H \,,\quad
  \h\psi  \mapsto \E(\h\psi):=\psi_e,                          &(4.7b)\cr}
$$
where $\psi_e$ is the component of $\h e$ in the expansion $(4.5)$.
It is easy to check that $\F(\psi)$ is indeed equivariant. We have
$$
\E\circ\F=\hbox{Id}{\Big\vert}_{\H}\,,\quad
\F\circ\E=\hbox{Id}{\Big\vert}_{\Hh}.
\eqno{(4.8)}
$$
The first equation is obvious, the second follows from $(4.6)$.
Hence $\E=\F^{-1}$. Moreover, we have
(an overbar over $\psi$ denotes complex conjugation)
$$
  \int_{\b Q}\l\F(\psi)(\b q)\vert\F(\phi)(\b q)\r\,d\b q
= \sum_{g,h}\int_{\b Q}\l\h g\vert\h h\r\,\b
   \psi(\b q g)\phi(\b q h)\,d\b q
= \int_{\b Q}\b \psi(\b q)\phi(\b q)\,d\b q,
\eqno{(4.9)}
$$
where we used $(2.31)$ in the last step. Hence $\F$ establishes
an unitary isomorphism between $\H$ and $\Hh$. The action $T$ of
$V_G$ on $\H$ can now be transferred to an action $\h T$ of $V_G$
on $\Hh$ via
$$
(\h v,\h\psi)\rightarrow{\h T}_{\h v}(\h\psi)\,,\quad
 {\h{T}}_{\h v}:=\F\circ T_{\h v}\circ\E
 \eqno{(4.10)}
$$
which yields, using $(4.7)$ and $(4.6)$,
$$\eqalignno{
\h{T}_{\h v}(\h\psi) &=\F\circ T_{\h v}(\psi_e)=
 \F\left(\sum_h v(h)\psi_e\circ R_{h^{-1}}\right)         &\cr
&=\sum_{g,h}\h gv(h)\psi_e\circ R_{h^{-1}}\circ R_{g}=
  \sum_{g,h}\h g v(h)\psi_e\circ R_{gh^{-1}}              &\cr
&=\sum_{f,h}\h f\cdot\h h\,v(h)\psi_f=\h \psi\cdot\h v\,. &(4.11)\cr}
$$
Hence $V_G$'s action on $\Hh$ just corresponds to pointwise right
multiplication. Note that
a pointwise left multiplication is not defined within $\Hh$ since the
resulting function would generally not be equivariant. But there is
such an action of left multiplications if one restricts to the centre
$V_G^c$.

Linear operators $\b O$ on $\H$ whose integral kernels satisfy $(1.12)$
($(1.13)$ for propagators) define linear Operators $\h O$ on $\Hh$
via $\h O:=\F\circ \b O\circ\E$ (for propagators:
$\h K(t';t):=\F\circ\b K(t';,t)\circ\E$).
As in $(4.11)$, we can easily derive the following explicit expressions
$$\eqalignno{
(\h O\psi)(\b q')&=
\int_{\b Q}\b O(\b q';\b q)\h \psi(\b q)\,d\b q,         &(4.12a)\cr
(\h K(t',t)\h\psi)(\b q')&=
\int_{\b Q}\b K({\b q}',t';\b q,t)\h\psi(\b q)\,d\b q\,, &(4.12b)\cr}
$$
which show that these operators just act componentwise on the functions
$\h\psi$, thus displaying manifestly the commutativity with the right
$V_G$-action:
$$
\h O\circ{\h{T}}_{\h v}={\h{T}}_{\h v}\circ\h O\,,\quad
\h K(t',t)\circ{\h{T}}_{\h v}={\h{T}}_{\h v}\circ \h K(t',t).
\eqno{(4.13)}
$$

Since the algebra $V_G$ now acts on the infinite-dimensional space $\Hh$
(or $\H$), we slightly adapt the basic notations from the previous section.
$B(\Hh)$ is the $C^*$-algebra of all bounded linear operators on $\Hh$
(similarly with $\H$). Through the implementation $(4.11)$, $V_G$ is
mapped linearly and anti-homomorphically (because of the
right-multiplication) onto a subalgebra of $B(\Hh)$,
which we call $\V_G$. It is not difficult to show that $\V_G$ is in
fact a von Neumann algebra.
A proof may be found in [GMN]\note{Although this
reference is primarily concerned with the symmetric group, the proof
given there works literally for any finite group.}. The actions of
$\h g$ or $\emu ij$ on $\Hh$ according to $(4.11)$ are denoted by the
linear operators ${\h{T}}_g$ or ${\h{T}}^{\mu}_{ij}$ respectively.
Accordingly, the linear operators corresponding to right
${\h e}^{\mu}_{i}$- and ${\h e}^{\mu}$-multiplications are projection
operators which we call ${\h{T}}^{\mu}_i$ and ${\h{T}}^{\mu}$.
They satisfy
$$
{\h{T}}^{\mu}_{ij} {\h{T}}^{\nu}_{kl}=
 \delta_{\mu\nu}\delta_{jk}{\h{T}}^{\mu}_{il},
\eqno{(4.14)}
$$
which follows directly from $(4.11)$ and $(2.14)$.

All the $H^*$-structural properties of $V_G$ are inherited by $\V_G$,
which makes it at the same time an $H^*$ and a von Neumann algebra.
{}From the definition of the scalar product $(4.4)$ it is obvious that
the two $*$-involutions so defined coincide. In particular,
${\h{T}}^{\mu}_i$ and ${\h{T}}^{\mu}$ are self-adjoint idempotents,
i.e., projection operators. The image of the subalgebras $V^{\mu}$,
$V^{\mu}_c$, $Z^{\mu}$ and $A$ will be called ${\V}^{\mu}$,
${\V}^{\mu}_c$, $\Z^{\mu}$ and $\A$ respectively. For any subset
$S\subset B(\Hh)$, $S'$ is the commutant which is in fact a von Neumann
algebra. $S''$ is called the von Neumann algebra generated by $S$,
which is equal to $S$ in case $S$ is already a von Neumann algebra.

Let us now look at the Hilbert space $\Hh$. We define the {\it algebra
of observables}, $\O$, by $\O:=(\V_G)'$. Its commutant then satisfies
$\O'=\V_G$. Further, the projection operators ${\h{T}}^{\mu}$ and
${\h{T}}^{\mu}_i$ define a split analogous to $(2.5)$
$$\eqalignno{
\Hh &= \bigoplus_{\mu=1}^m  {\Hh}^{\mu},                &(4.15a)\cr
{\Hh}^{\mu} &= \bigoplus_{i=1}^{n_{\mu}}{\Hh}^{\mu}_i,  &(4.15b)\cr}
$$
where ${\Hh}^{\mu}={\h{T}}^{\mu}\Hh$ and
${\Hh}^{\mu}_i={\h{T}}^{\mu}_i\Hh$. The functions in these Hilbert
spaces are just given by the $V^{\mu}$- and $L^{\mu}(e_i)$-valued
functions in $\Hh$ respectively.

The second split of course inherits the non-uniqueness from $(2.5b)$.
Under a redefinition $(2.28)$ we just have to analogously conjugate
the matrix ${\h T}^{\mu}$ by $U^{\mu}$. For example, given a
normalized $a=\sum_ia_ie_i\in C^{n_{\mu}}$, we can choose it as the
first basis vector of a new basis $e'_i=\sum_j U^{\mu}_{ij}e_j$ with
$a_i=U^{\mu}_{1i}$. The projection operator onto $L^{\mu}(a)$-valued
functions is then given by
${\h T}^{\mu}(a):=\sum_{i,j}a_i{\b a}_{j}{\h T}^{\mu}_{ij}$.

The operators and propagators in $(4.12)$ now
project into each subspace:
$$\eqalignno{
 {\h O}^{\mu}:=&{\h{T}}^{\mu}\circ \h O\circ{\h{T}}^{\mu},
&(4.16a)\cr
 {\h O}^{\mu}_i:=&{\h{T}}^{\mu}_i\circ\h O\circ{\h{T}}^{\mu}_i,
&(4.16b)\cr}
$$
where, since $O\in \O$, the left projection operators are not
really necessary. The analogous formulae hold for the propagator.
It is then obvious that the projected propagators in ${\Hh}^{\mu}_i$
satisfy the standard combination rule:
$$
\int_{\b Q} {\h K}^{\mu}_i({\b q}',t';{\b q}'',t'')\,
            {\h K}^{\mu}_i({\b q}'',t'';\b q,t)\,d{\b q}''
           ={\h K}^{\mu}_i({\b q}',t';\b q,t),
\eqno{(4.17)}
$$
and the analogous relations for ${\h K}^{\mu}$ by summing over $i$.
The latter ones are then exactly the non-abelian versions of $(1.19)$,
only expressed in terms of $\Hh$ rather than $\H$.
Here, in the non-abelian case, we have a finer splitting due
to the $n_{\mu}$-fold multiplicity (labeled by the index $i$) of
the $\mu$-th representation.

Clearly, everything said for $\Hh$ can be easily translated to $\H$
using the unitary equivalence $(4.7)$. For example, the projection maps
$T^{\mu}_i$, $T^{\mu}(a)$ and the projected integral kernels of
propagators and operators take the form
$$\eqalignno{
& {T}^{\mu}_i\psi={n_{\mu}\over n}\sum_{g\in G}
  D^{\mu}_{ii}(g)\,\psi\circ R_g  \,,                   &(4.18a)\cr
& {\b K}^{\mu}_i({\b q}',t';\b q,t)=
  {n_{\mu}\over n}\sum_{g\in G}
  D^{\mu}_{ii}(g)\,{\b K}({\b q}' g,t';\b q,t) \,,      &(4.19)\cr
& {\b O}^{\mu}_i({\b q}',\b q)=
  {n_{\mu}\over n}\sum_{g\in G}D^{\mu}_{ii}(g)\,
                 \b O({\b q}' g,\b q)\,,                &(4.20)\cr}
$$
and equivalently (by summing these expressions over $i$) for $T^{\mu}$
and $O^{\mu}$. As explained above, the most general expression for
a projector is given for some normalized $a\in C^{n_{\mu}}$ by
$$
{T}^{\mu}(a)\psi={n_{\mu}\over n}\sum_{g,i,j}a_i{\b a}_j
D^{\mu}_{ij}(g)\,\psi\circ R_g\,.
\eqno{(4.18b)}
$$
In the same way $(4.19)$ and $(4.20)$ can be written in terms of $A$.
All these expressions
form the non-abelian generalization of $(1.15)$, $(1.18)$ and $(1.20)$.
An application of $(4.18)$ appears in appendix B. As already mentioned,
$(4.17)$ hold literally for $\b K$ instead of $\h K$. In the present
setting this is obvious from construction, though it can of course also
be verified explicitly from $(4.19)$ and $(2.8a)$. For many of the
general aspects we consider here it is however more convenient to
work with $\Hh$ rather than $\H$.

Coming back to the definition of observables on $\Hh$, they do not only
include those of the form $(4.12a)$, but also right multiplications
with elements in the centre $\V_G^c$ of $\V_G$, that is, the algebra
generated by $\{{\h T}^1,\dots, {\h T}^m\}$. We now state the main
structural properties of the pair $(\Hh,\O)$ in the following

\proclaim
Theorem. (i) $\O$ is completely reducible. The subspaces ${\Hh}^{\mu}_i$
             are minimal invariant relative to $\O$.
        (ii) A subspace ${\Hh}'\subset \Hh$ reduces $\O$ and $\V_G$,
             iff ${\Hh}'=\bigoplus_{\mu\in J}{\Hh}^{\mu}$, where
             $J$ is a subset of $\{1,\dots, m\}$.
       (iii) A minimal invariant subspace ${\Hh}^{\mu}_i$
              reduces $\Z^{\mu}$. It reduces $\V_G$, iff the
              $\mu$-th representation is abelian.\par

\noindent
{\bf Proof.}
(i)   Suppose ${\Hh}^{\mu}_i={\h T}^{\mu}_i\Hh$ were reducible under
      $\O$. Then there existed two orthogonal self-adjoint idempotents
      ${\h S}^{\mu}_i$ and ${\h P}^{\mu}_i$ with
      ${\h T}^{\mu}_i={\h S}^{\mu}_i+{\h P}^{\mu}_i$ and
      ${\h S}^{\mu}_i,{\h P}^{\mu}_i\in\O'=\V_G$ (since $\V_G$ is von
      Neumann). But this cannot be, since from the structural
      properties of $V_G$ we know that the $T^{\mu}_i$ are already
      minimal idempotents.
(ii)  From (i) we have ${\Hh}^{\mu}_i={\h T}^{\mu}_i\Hh\subset {\Hh}'$
      for some pair $\mu,i$. By hypothesis
      ${\h T}^{\mu}_{ki}{\Hh}^{\mu}_i\subset{\Hh}'$ for any $k$,
      and these subspaces are clearly non null.
      Using $(4.14)$, the left side can be rewritten as
      ${\h T}^{\mu}_{ki}{\h T}^{\mu}_{i}\Hh=
      {\h T}^{\mu}_{k}{\h T}^{\mu}_{ki}\Hh \subset{\Hh}^{\mu}_k$.
      Hence there is a non trivial intersection
      ${\Hh}'\cap{\Hh}^{\mu}_k$ $\forall k$, which by (i) implies
      $\bigoplus_k{\Hh}^{\mu}_k={\Hh}^{\mu}\subset{\Hh}'$.
(iii) It reduces $\Z^{\mu}$ since it commutes with $\V_G$.
      To reduce $\V_G$ it is clear from (i) and (ii) that $\mu$ must
      be such that the range of $i$ is only $1$, i.e., $n_{\mu}=1$.
      But this is the case iff the $\mu$-th representation is
      abelian~$\bullet$

To conform with the Jauch-Wightman requirement, we proceed exactly as in
the  previous section. For each $\mu$ we truncate the Hilbert space
so as to contain only one summand in $(4.15b)$, say
${\Hh}^{\mu}_1=:{\Hh}^{\mu}_{\rm tr}$, and obtain
$$
{\Hh}_{\rm tr}=\bigoplus_{\mu=1}^m {\Hh}^{\mu}_{\rm tr}.
\eqno{(4.21)}
$$
Accordingly, the algebra of observables can now be written as
$$
\O=\bigoplus_{\mu=1}^m B({\Hh}^{\mu}_{\rm tr})
\eqno{(4.22)}
$$
which is the general form of the algebra of observables in any theory
with standard\note{Superselection rules are said to be standard, if
they are commutative, and in addition the linear span of the pure states
lies dense in the Hilbert space. The latter condition is known as the
condition of {\it discrete} superselection rules, since it ensures the
decomposability into a discrete direct sum (rather than a direct integral)
of irreducible representations (possibly with multiplicities) of the
algebra of observables. In short: commutative + discrete = standard.
For the nomenclature, see e.g. [BLOT]. However, the
condition of discreteness is often violated even in standard quantum
mechanics. For example, the mass superselection rule in Galilean
invariant quantum mechanics is continuous, since each mass value defines
a separate sector (compare [Gi]).} superselection rules [BLOT].
Its representation on ${\Hh}_{\rm tr}$ is phenomenologically equivalent
to its representation on $\Hh$, but pure states are now in bijective
correspondence with rays in the set
$$
\bigcup_{\mu=1}^m {\Hh}^{\mu}_{\rm tr}  \,.
\eqno{(4.23)}
$$
In each sector $\Hh^{\mu}_{\rm tr}$ the group $C^{\mu}$ is still acting.
All these features are just like in the finite-dimensional model.

It is important to note that the definition $\O={\V'}_G$
yields a richer set of observables than those coming from quantizing
functions on the non-redundant classical phase space $T^*(Q)$. This is
obvious from $(4.12a)$, since the operators do not act on the ``internal''
vector space. But since $\O$ acts irreducibly in the sectors
${\Hh}_{\rm tr}^{\mu}$, as asserted by the theorem above, there must be
additional observables for the non-abelian sectors [So][Ba2].
For example, for non-abelian sectors, any localization on the true
configuration space $Q$ still does not specify in any way the direction
of the ``internal'' vector. In order to fix it, additional observables
must be employed. These observables cannot simply be given by
pointwise left $V_G$-multiplication, for, as we have seen above,
only elements of $Z^{\mu}$ act on ${\Hh}_{\rm tr}^{\mu}$, where they
are necessarily proportional to the identity operator.
However, if we first apply some localization to the system in
configuration space, we can indeed define observables acting on the
``internal'' space. Let us explain this in more detail.

Let $U\subset Q$ be a closed connected\note{Connectedness is not a
relevant requirement and may without gain or loss just as well be
dropped. It does simplify the argument however.} subset and
$\b U\subset \b Q$ a connected covering set. We call $U$ admissible if
$\b U\cap\b Ug=\empty$ $\forall g\not =e$. Here, $\b Ug$ is the right
translation of $\b U$ by $g$. We call $\h\psi$ $U$-localized, iff its
support is contained in the interior of $\bigcup_g \b Ug$. This defines
a linear subspace $\Hh_U$ of $U$-localized states. Note that the
variety of admissible subsets $U$  is very big. In particular they
contain all contractible subsets of $Q$. Also, the set may be chosen
such as to leave a complement with arbitrarily small volume. However,
physically it might be more relevant to think of the admissible sets as
being rather small
portions of $Q$ on which realistic ``filters'' project.
Any localized state is completely determined by its restriction to $\b U$.
Let $\chi_{\b U}$ be the characteristic function of $\b U$, and
$\chi_{\b Ug}=\chi_{\b U}\circ R_{g^{-1}}$ those of the translated sets.
We set ${\h\psi}_{\b Ug}=\chi_{\b Ug}\h\psi$. Equivariance $(4.3)$ implies
that ${\h\psi}_{\b Ug}={\h g}^{-1}\cdot\h\psi_{\b U}\circ R_{g^{-1}}$.
A projection operator $P_U:\,\Hh\rightarrow\Hh_U$ is then given by
$$
P_U(\h\psi):= \sum_{g\in G}{\h\psi}_{\b Ug}
            = \sum_{g\in G}\h g\cdot{\h\psi}_{\b U}\circ R_g\,,
\eqno{(4.24)}
$$
and functions $\h\psi\in \Hh_U$ are determined by their restriction
$\h\psi_{\b U}$. Since $\tau:\b U\rightarrow U$ is a diffeomorphism,
we can also use the pullback $\h\psi_U:=\h\psi_{\b U}\circ\tau^{-1}$
on $Q$. Now, on $\Hh_U$ we can define a left $V_G$-action as follows:
for $\h v=\sum_h v(h)\h h$ we set
$$
(\h v,\h\psi)\mapsto \gamma_{\h v}(\h\psi)
:=\sum_{h\in G}v(h)\sum_{g\in G}\h g\cdot \h h\cdot\h\psi_{\b U}
\circ R_g\,.
\eqno{(4.25)}
$$
It is easily seen that this is indeed a map from $\Hh_U$ to $\Hh_U$,
in particular, $\gamma_{\h v}(\h\psi)$ is equivariant. Moreover, this
action commutes with $\V_G$ since it clearly commutes with right $V_G$
multiplications. It therefore also defines an action on $U$-localized
states in $\Hh_{\rm tr}$ and each sector $\Hh_{\rm tr}^{\mu}$ separately.
For general (i.e. non localized) states, observables may be defined by
first projecting with $P_U$ on any admissible $U$ and then applying
$\gamma_{\h v}$:
$$
O_{\h v}:=\gamma_{\h v}\circ P_U.
\eqno{(4.26)}
$$
One easily verifies that this is a self-adjoint operator iff
$\h v={\h v}^*$. On the local representative $\h\psi_U$ on $U\subset Q$
this just corresponds to left $\h v$-multiplication.  This construction
seems to implement some ideas presented in [So][Ba2].
It would be interesting to explicitly construct and interpret these
observables in simple models.

Everything we have said could be rephrased in terms of the possibly more
familiar language of vector bundles over $Q$. Sections of this bundle
could be represented by locally defined  functions like $\h\psi_U$.
This is explained in detail in the following appendix A. We have
deliberately avoided this language in order to always deal with globally
defined functions (on $\b Q$). In particular, the left $G$-action
defined on localized states through $(4.25)$ should not be confused with
gauge transformations. We refer to appendix A for more details.

Finally we make a few comments on the implementation of symmetries.
The issue is whether we can always assume the symmetries to respect
the sector structur, that is, whether symmetries that initially act
on $\H$ are reduced by the subspaces ${\H}^{\mu}$ and
${\H}^{\mu}(a)=T^{\mu}(a)\H$. If the unitary symmetry operators
commute with ${\V}_G$, i.e., are elements in $\O$, all subspaces
that reduce $\O$ also reduce the symmetry group and there is no
problem with its implementation in the sectors.
This is the case for continuous groups whose generators should
correspond to physical quantities and therefore commute with
${\V}_G$ (in the sense of section VIII.5 in [RS]). But there are
discrete symmetries which do not commute with ${\V}_G$, like
time-reversal. In fact, if the complex conjugate representation,
${\b D}^{\mu}$, of $D^{\mu}$ is not equivalent to $D^{\mu}$, i.e.,
${\b D}^{\mu}=D^{\lambda}$, $\lambda\not =\mu$, complex conjugation
will connect two different sectors. The operation of time-reversal
is therefore not implementable in these sectors. They are said to
`break' time-reversal invariance. For abelian sectors this is the
case iff the representation is not real [Sch]. Conversely, if we have
${\b D}^{\mu}=U^{\dagger}D^{\mu}U$, then $(4.18b)$ shows
$\overline{T^{\mu}(a)\psi}=T^{\mu}(\overline{Aa}){\b \psi}$.
Since the truncated Hilbert space ${\H}^{\mu}_{\rm tr}$ can be
identified with any of the ${\H}^{\mu}(a)$, which are mutually
isomorphic in a natural way, we can use this isomorphism to map back
${\H}^{\mu}(\overline{Aa})$ to ${\H}^{\mu}(a)$ and thus define the
operator of time-reversal on ${\H}^{\mu}_{\rm tr}$. We avoid to write
down the details at this point which immediately follow from our
general discussion in section 3. We conclude that the $\mu$-th sector
breakes time-reversal invariance, iff the representation $D^{\mu}$
is inequivalent to its complex conjugate. (For a general criterion
see chapter 5-5 in [Ha].)

\beginsection{Appendix A}

In this appendix we recall some basic features of principal bundles
and their associated vector bundles as applied to the universal
covering space. As already stated in section 1, the universal covering
space $\b Q$ is the total space of a principal fibre bundle with
structure group $G\cong \pi_1(\b Q,\b q)$, base $Q$ and projection
$\tau:\b Q\rightarrow Q$. $G$ acts on $\b Q$ via right multiplications:
$R_g(\b q)=\b qg$, so that $\tau(\b qg)=\tau(\b q)$ for all $g\in G$.
The action is
transitive on each fibre $\tau^{-1}(q)$. Discreteness of the
fibres implies that $\tau_*:\,T_{\b q}(Q)\rightarrow T_{q}(Q)$ and
${R_g}_*:\,T_{\b q}(\b Q)\rightarrow T_{\b qg}(\b Q)$ are both
isomorphisms. We can thus trivially regard $T_{\b q}(\b Q)$ as its own
horizontal subspace. This defines a naturally given connection as
follows: given a loop, $\gamma:\,[0,1]\rightarrow Q$, based at
$\gamma(0)=\gamma(1)=q$, we have for each $\b q\in\tau^{-1}(q)$ a
unique (horizontal) lift, $\b\gamma:\,[0,1]\rightarrow\b Q$, such that
$\b\gamma(0)=\b q$ and $\b\gamma(1)=\b qg$ for some uniquely determined
$g\in G$. Since $G$ is discrete, $g$ depends only on the homotopy class
$[\gamma]\in\pi_1(Q,q)$. This defines the family of maps
$$
I_q:\,\pi_1(Q,q) \rightarrow G\,,\quad
      [\gamma]   \mapsto g=:I_q([\gamma])\,.
\eqno{(A.1)}
$$
Choosing a different point, ${\b q}'=\b qh\in\tau^{-1}(q)$, the lift of
$\gamma$ starting at ${\b q}'$ is now given by
${\b\gamma}'=R_h\circ\b\gamma$, which ends at
${\b\gamma}'(1)=\b\gamma(1)h=\b qgh={\b q}'h^{-1}gh$, so that
$$
I_{\b qh}={\hbox{Ad}}(h^{-1})\circ I_{\b q}.
\eqno{(A.2)}
$$
Moreover, $I_{\b q}([\gamma_1][\gamma_2])$ is defined by lifting
$[\gamma_1\gamma_2]$\note{We adopt the standard convention that
products of paths are read from the left, that is, $\gamma_1\gamma_2$
is $\gamma_1$ followed by $\gamma_2$. If we read it from right to
left, like maps, the $I_{\b q}$ would be isomorphisms in $(A.3)$}:
lifting $\gamma_1$ takes one from $\b q$ to
${\b q}'=\b qI_{\b q}([\gamma_1])$, and the lift of $\gamma_2$ then
from ${\b q}'$ to ${\b q}'I_{{\b q}'}([\gamma_2])$, which, using $(A.2)$,
is equal to $\b qI_{\b q}([\gamma_2])I_{\b q}([\gamma_1])$. Hence
each $I_{\b q}$ defines an anti-isomorphism:
$$
I_{\b q}([\gamma_1][\gamma_2])=
I_{\b q}([\gamma_2])I_{\b q}([\gamma_1]).
\eqno{(A.3)}
$$
As already mentioned in section 1, there is generally no natural
isomorphism between the fundamental groups at different points
and $G$. For example, looping the basepoint along $\gamma$ results
in a conjugation with $[\gamma]$ (see e.g. [St], paragraph 16).
Identifications with an abstract group $G$ are therefore only defined
up to inner automorphisms. This at least provides a natural
identification of conjugacy classes of all $\pi_1(Q,q)$ with those of
$G$. Unless one refers to a basepoint, it generally does not make sense
to talk about {\it the} fundamental group, or a specific element thereof.
But it does make sense to speak of {\it a particular} conjugacy class.
For example, if $g\in G_c$ (the centre), it makes sense to call it a
{\it particular} element of {\it the} fundamental group. If restricted
to the centre, the maps $I_{\b qh}$ are independent of $h$, as $(A.2)$
shows. Right multiplication by the central element $g$ might therefore
be interpreted as ``parallelly transporting each element of $\b Q$ along
the loop $g$''. For elements not in the centre this notion is not defined.

Since $\b Q$ is a principal bundle, we can also apply the concept
of gauge transformations. These are given by diffeomorphisms
$F:\,\b Q\rightarrow\b Q$ such that $F\circ R_g=R_g\circ F$ (bundle
automorphisms), and $\tau\circ F=\tau$ (projecting to the identity on $Q$).
It is easy to see that any such function $F$ can be written in the form
$F(\b q)=\b qf(\b q)$, with a uniquely determined smooth function
$f:\,\b Q\rightarrow G$ satisfying $f\circ R_g={\rm Ad}(g^{-1})\circ f$.
In that sense gauge transformations uniquely correspond to Ad-equivariant,
$G$-valued functions on $\b Q$. The composition $F=F_1\circ F_2$ corresponds
to the function $f=f_1f_2$, where juxtaposition on the right hand side
means pointwise multiplication in $G$.
However, since in our case $G$ is discrete, the $G$-valued function $f$
must be constant. Ad-equivariance then implies that it must assume values
in the centre $G_c$ of $G$. The group of gauge transformations is therefore
given by right $G_c$-multiplications. {\it In particular, the group of gauge
transformations does not contain the gauge group if $G$ is non-abelian.}

\subsection{Vector Bundles}

Let $V$ be a (complex) vector space and $D^{\mu}$ an (irreducible)
representation of $G$ on $V$. We can associate to the principal bundle
$\b Q$ a vector bundle, $E(Q,V,G,\mu)$, with base $Q$, fibre $V$,
structure group $G$, and total space $E$:
$$\eqalign{
E= &  \{\b Q\times V\}/\sim \,,    \cr
\hbox{where}\quad (\b q,v)\sim(\b p,w) & \Leftrightarrow\exists
g\in G\,/\,\b p=\b qg,\ w=D^{\mu}(g^{-1})v.\cr}
\eqno{(A.4)}
$$
We denote the equivalence class of $(\b q,v)$ by $[\b q,v]$, and have the
inherited projection map $\tau_E:\,E\rightarrow Q$,
$\tau_E([\b q,v]):=\tau(\b q)=q$. Parallel transportation of
$[\b q,v]\in\tau_E^{-1}(q)$ along a curve $\gamma$ in $Q$ from
$\gamma(0)=q$ to $\gamma(1)=p$ is defined as follows: Take the
horizontal lift $\b\gamma$ of $\gamma$ on $\b Q$, such that
$\b\gamma(0)=\b q$. This defines a curve $\t\gamma$ in $E$ via
$\t\gamma:=[\b\gamma,v]$. Its end point,
$\t\gamma(1)=[\b\gamma(1),v]\in\tau_E^{-1}(p)$, then defines the parallel
transport of $[\b q,v]$. In particular, if $\gamma$ is a loop at
$q\in Q$, we have, using the notation above,
$\t\gamma(1)=[\b\gamma(1),v]=[\b qg,v]=
[\b q,D^{\mu}(g)v]=[\b q,D^{\mu}(I_{\b q}([\gamma]))v]$.
This defines a family of holonomy maps
$H_q$:
$$\eqalign{
H_{\b q}:\ \pi_1(Q,q) & \rightarrow\hbox{End}(\tau_E^{-1}(q)), \cr
H_{\b q}([\gamma])([\b q,v]) &
:=[\b q,D^{\mu}(I_{\b q}([\gamma]))v], \cr}
\eqno{(A.5)}
$$
which is an anti-homomorphism, due to $(A.3)$. Note that the right action
$R_g$ on $\b Q$ does generally not define an action -- hypothetically
denoted by $\gamma_g$ -- on $E$, since in this case
$\gamma_g([\b q,v]):=[\b qg,v]=[\b q,D^{\mu}(g)(v)]$
should equal $\gamma_g([\b qh,D^{\mu}(h^{-1})v]=[\b qhg,D^{\mu}(h^{-1})v]
=[\b q, D^{\mu}(hgh^{-1})v]$ for all $h$. This is the case iff
$g\in C^{\mu}$, where
$C^{\mu}=\{h\in G\,/\,D^{\mu}(hg)=D^{\mu}(gh)\,\forall g\in G\}$;
in words, $C^{\mu}$ is the largest subgroup of $G$ which under
$D^{\mu}$ maps into the centre of $D^{\mu}(G)$.
One also easily verifies that $D^{\mu}(g)=D^{\mu}(g_G)$
$\forall g\in C^{\mu}$, where $g_G$ is the conjugacy class of $g$.
Thus, although there is generally no action of $G$ on $E$,
there is such an action of $C^{\mu}$:
$$
\gamma_g:\ [\b q,v]\rightarrow[\b qg,v]=[\b q,D^{\mu}(g)v]\
        \forall g\in C^{\mu}\,.
\eqno{(A.6)}
$$
Allowing some abuse of language, we may say that this
corresponds to a parallel transportation along a loop at $q$
representing $g\in C^{\mu}$. As explained above we should actually
refer to the whole class $g_G$, but the ambiguity in assigning a
particular member of $g_G$ to each $\pi_1(Q,q)$ is projected out
due to $D^{\mu}$ being constant on $g_G$.

Finally, given a cross section $\sigma$ in $E$, we can define an
action of $C^{\mu}$ on $\sigma$. To see this explicitly, recall
that for each section $\sigma$ there is a unique $D^{\mu}$-equivariant
function $\b\sigma$ on $\b Q$:
$$
\b\sigma:\, \b Q\rightarrow V\,,\quad\b\sigma\circ R_g=
D^{\mu}(g^{-1})\,\b\sigma,
\eqno{(A.7)}
$$
defined implicitly by
$$
\sigma(q)=[\b q,\b\sigma(\b q)].
\eqno{(A.8)}
$$
Alternatively, sections in $E$ can be described locally on $Q$. Given
a local section $\lambda:\, U\rightarrow \b Q$ on an open subset
$U\subset Q$, we have the locally defined $V$-valued function on $U$:
$$
\sigma_{\lambda}:\, U\rightarrow V\,,\quad\sigma_{\lambda}:=
\b\sigma\circ\lambda\,.
\eqno{(A.9)}
$$
On $U$ it satisfies $[\lambda(q),\sigma_{\lambda}(q)]=\sigma(q)$.
Any other local section, $\lambda':\,U\rightarrow \b Q$, is necessarily
of the form $\lambda'=R_h\circ\lambda $ for some $h\in G$.
We then have, using $(A.9)$,
$$
\sigma_{\lambda'}=D^{\mu}(h^{-1})\sigma_{\lambda}.
\eqno{(A.10)}
$$

Now, an action (also denoted by $\gamma_g$) of $C^{\mu}$ on the
section $\sigma$ is just given by the obvious choice
$$
(\gamma_g\sigma)(q):=
[\b qg,\b\sigma(\b q)]=[\b q,D^{\mu}(g)\b\sigma(\b q)].
\eqno{(A.11)}
$$
Equivalently, expressed in terms of $\b\sigma$ or the local
representative $\sigma_{\lambda}$, we have
$$\eqalignno{
(\gamma_g\b \sigma)(\b q) &= \b\sigma(\b qg^{-1})=
                        D^{\mu}(g)\b\sigma(\b q), &(A.12)\cr
(\gamma_g\sigma_{\lambda})(q) &= D^{\mu}(g)\sigma_{\lambda}(q). &(A.13)\cr}
$$
As above, we could -- again with some abuses of language -- say that
$\gamma_g\sigma$ is the result of ``parallelly transporting the section
$\sigma$ along a loop  representing (the class of) $g$ in the
fundamental group''.

Quite generally, in gauge theory one cannot use the local formula
$(A.13)$ as definition of an action of the gauge group. {\it
The gauge group simply does not act on the space of sections in
the general case.} However, in special circumstances meaning can be
given to a definition in the form $(A.13)$ in the following  way: Let
$U\subset Q$ and $\lambda$ as before and $\Gamma_U$ the linear space
of sections $\sigma:\,Q\rightarrow E$ whose support is contained in $U$.
We now use the distinguished section $\lambda$ to {\it define} a
$G$-action on $\Gamma_U$ via $(A.13)$. With respect to a different
section $\lambda'=R_h\circ\lambda$ the so defined action reads
$$
\gamma_g\sigma_{\lambda'} = D^{\mu}(h^{-1}gh)\sigma_{\lambda'}\,.
\eqno{(A.14)}
$$
The best way to see that this defines indeed a $G$-action on $\Gamma_U$ is
to express it in terms of the globally defined (on $\b Q$) equivariant
functions $\b\sigma$ and check that the result is again equivariant.
To do this, let $\lambda(U)=\b U\subset \b Q$ and recall that the
restriction $\b\sigma\big\vert_{\b U}$ determines all other restrictions
$\b\sigma\big\vert_{\b Ug}$ by equivariance. From $(A.7)$ one has
$\b\sigma\big\vert_{\b Ug}=D^{\mu}(g^{-1})\b\sigma\big\vert_{\b U}
\circ R_{g^{-1}}$.
That $\sigma$ is in $\Gamma_U$ means here that $\b\sigma$ has support in
$\bigcup_{g\in G}{\b U}g\subset \b Q$. We define the function
$\b\sigma_{\b U}$ on $\b Q$ to equal the restriction
$\b\sigma\big\vert_{\b U}$ within $\b U$ and be identically zero
otherwise. We can then express $\b\sigma$ as a sum of terms
with disjoint support:
$$
\b\sigma=\sum_{h\in G} D^{\mu}(h){\b\sigma}_{\b U}\circ R_h\,.
\eqno{(A.15)}
$$
Since $\b\sigma_{\b U}$ is essentially $\sigma_{\lambda}$, the
action defined by $(A.13)$ now reads
$$
\gamma_g\b\sigma=\sum_{h\in G} D^{\mu}(hg){\b\sigma}_{\b U}\circ R_h,
\eqno{(A.16)}
$$
which is again equivariant. What happened here is that in the support
component $\b Uh$ the function $\b\sigma$ is multiplied with
$D^{\mu}(h^{-1}gh)$, as required by $(A.14)$. Here the additional
conjugation is necessary for the result to be equivariant. This
definition would be contradictory if the support were not inside the
disjoint regions $\b Ug$. This is the reason why we had to restrict to
$\Gamma_U$

There is a certain danger to misunderstand this construction in the
following way: the restriction to $\Gamma_U$ effectively truncates
the principal bundle $\b Q$ to $\tau^{-1}(U)$ which is itself a
trivial bundle. Given a distinguished section $\lambda$ in this
truncated bundle there is a induced trivialization
$\tau^{-1}(U)\rightarrow U\times G$ given by\note{Any element in
$\tau^{-1}(U)$ can be uniquely written as $\lambda (q)h$.}
$\lambda(q)h\mapsto (q,h)$. Then there is a left
action $\gamma_g$ of $G$ defined by $\gamma_g(q,h)=(q,gh)$ or
$\gamma(\lambda(q)h)=\lambda(q)gh$. This clearly defines a gauge
transformation $F$ of $\tau^{-1}(U)$ which is easily seen to
induce the action $\gamma_g$ on sections. This suggests the incorrect
conclusion that our action $\gamma_g$ is really nothing but a gauge
transformation. The point is that the map $F$ will not extend from
$\tau^{-1}(U)$ to $\b Q$, so that we are {\it not} dealing with a gauge
transformation on $\b Q$ or $E$.

\beginsection{Appendix B}

A simple mechanical system with finite non-abelian fundamental group is
the non-symmetric rotor. It serves, for example, as a dynamical model
for the collective rotational degrees of freedom of deformed nuclei [BM].
In this appendix we explicitly construct the sectors by applying formula
$(4.18)$ to the standard basis functions. This leads
precisely to the known symmetry classification of collective rotational
modes of nuclei but interprets it in the present formalism.
In particular, the only sector for odd-A nuclei corresponds to
non-abelian representations of the fundamental group. This relevant
sector would have been lost if one restricted to abelian
representations. This example therefore serves to illustrate our
discussion at the end of section 3.

The different configurations for the non-symmetric rotor are easily
visualized as the different orientations of a solid ellipsoid with
pairwise different major axes. Its symmetries are generated by
$\pi$-rotations about any two of the three major axes and form the
group $Z_2\times Z_2$. The configuration space is thus given by
$SO(3)/Z_2\times Z_2$, but it is more conveniently represented
by $SU(2)/D_8^*$, where $D_8^*$ is the preimage of $Z_2\times Z_2$
under the 2-1 projection $SU(2)\rightarrow SO(3)$. $D_8^*$ is called
the binary dihedral group of order eight and is conveniently defined
using unit quaternions:
$D_8^*=\{\pm 1,\pm {\rm i},\pm {\rm j},\pm {\rm k}\}$, where
${\rm i}^2={\rm j}^2={\rm k}^2=-1$, ${\rm i}{\rm j}={\rm k}$ and
cyclic. The configuration space is thus defined by $Q:=SU(2)/D_8^*$.
Since $SU(2)\cong S^3$ is simply connected, we have $\b Q=S^3$ and
$\pi_1(Q)\cong D_8^*$.

We consider the Hilbert space $\H=L^2(S^3,d\b q)$ where $d\b q$ is the
measure induced by the kinetic energy metric of the rotor. Such a
metric is invariant under left $SU(2)$ and right $D_8^*$
multiplications\note{We adopt the standard convention that left
multiplications correspond to rotations in the space-fixed and
right multiplications to rotations in the body-fixed frame. The
identifications under $D_8^*$ are therefore done using the right
multiplications.}, and so is the measure $d\b q$. Let
$\{R_{MN}^{\Lambda}\}$, $2\Lambda=0,1,2..$, denote the
representation matrices for $SU(2)$. We use the standard convention
to label the $2\Lambda+1$ values for the indices $M,N,..$ by
$\{-\Lambda,-\Lambda+1,\dots,\Lambda\}$. We can now expand each
$\psi\in\H$
in the form\note{In order to properly normalize the basis functions we
would have to multiply each $R^{\Lambda}_{MN}$ with a factor
proportional to $\sqrt{I_1I_2I_3(2\Lambda+1)/16\pi^2}$. The moments
of inertia, $I_i$, appear because they need to be cancelled from the
measure derived from the kinetic energy metric.}:
$$
\psi=\sum_{M,N,\Lambda}
     C^{\Lambda}_{MN}{R}^{\Lambda}_{MN}\,.
\eqno{(B.1)}
$$

$D_8^*$ has four one-dimensional irreducible representations,
$D^0,D^1,D^2,D^3$, and one two-dimensional one, $D^4$. The one-dimensional
representations are labelled by three $\{-1,1\}$-valued numbers,
$(r_1,r_2,r_3)$, where $D^{\mu}(\pm {\rm i})=r_1$,
$D^{\mu}(\pm {\rm j})=r_2$, and $D^{\mu}(\pm {\rm k})=r_3$, so that:
$$
(r_1,r_2,r_3)=\cases{ (1,1,1)   & for $\mu=0$, \cr
                      (1,-1,-1) & for $\mu=1$, \cr
                      (-1,1,-1) & for $\mu=2$, \cr
                      (-1,-1,1) & for $\mu=3$. \cr}
\eqno{(B.2)}
$$
One sees that it is in fact sufficient to uniquely characterize a
one-dimensional representation by two of the three $r_i$'s. We shall
take  $r_2$ and $r_3$. The two-dimensional representation, $D^4$, can
be defined using the standard Pauli-matrices $\{\tau_1,\tau_2,\tau_3\}$:
$$\eqalign{
D^4(\pm 1)       & =\pm i{\bf 1}, \cr
D^4(\pm {\rm i}) & =\mp i\tau_1,  \cr
D^4(\pm {\rm j}) & =\mp i\tau_2,  \cr
D^4(\pm {\rm k}) & =\mp i\tau_3.  \cr}
\eqno{(B.3)}
$$

Using standard results from finite group
theory\note{Here we just use the formula
$n^{\Lambda}_{\mu}={1\over 8}\sum_{g\in D_8^*}{\bar\chi}^{\mu}(g)
\chi^{\Lambda}(g)$ for the number of times $D^{\mu}$ is contained
in $R^{\Lambda}$. $\chi^{\mu}$ and $\chi^{\Lambda}(g)=
\sin((\Lambda+{1\over 2})\alpha)/\sin({\alpha\over 2})$ are the
characters of $D^{\mu}(g)$ and $R^{\Lambda}(g)$ respectively, and
$\alpha$ is the rotation angle of $g$.}
one easily finds that
for even $\Lambda$, $D^0$ occurs $({\Lambda\over 2}+1)$-times and
$D^{1,2,3}$ each ${\Lambda\over 2}$-times, whereas for odd $\Lambda$
$D^0$ occurs ${\Lambda -1\over 2}$-times and $D^{1,2,3}$ each
${\Lambda+1\over 2}$-times. $D^4$ is of course not contained in
representations with integer $\Lambda$. Conversely, for
$\Lambda={{\rm odd}\over 2}$ only the two-dimensional representation
$D^4$ occurs, namely $(\Lambda+{1\over 2})$-times.
All representations are equivalent to their complex conjugates.
This is trivial for the one-dimensional ones, which are real, and
for $D^4$ we have ${\b D}^4=U^{\dagger}D^4U$ with $U=i\tau_2$.

We are interested in the projection operators $T^{\mu}(a)$, written
down in $(4.18b)$. We first deal with the abelian cases $\mu=0,1,2,3$. Here
$R^{\Lambda}({\rm j}{\rm k})=R^{\Lambda}({\rm k}{\rm j})$. It is
convenient to introduce the four projector matrices:
$$\eqalign{
P^{\Lambda}_{r_2}({\rm j}) &=\shalf\left[{\bf 1}+r_2R^{\Lambda}({\rm j})
\right],\cr
P^{\Lambda}_{r_3}({\rm k}) &=\shalf\left[{\bf 1}+r_3R^{\Lambda}({\rm k})
\right].\cr}
\eqno{(B.4)}
$$
We easily find
$$
T^{\mu}R^{\Lambda}=R^{\Lambda} P^{\Lambda}_{r_2}({\rm j})
                               P^{\Lambda}_{r_3}({\rm  k})\,,
\eqno{(B.5)}
$$
where the right hand side is understood as multiplication of the
matrix-valued function $R^{\Lambda}$ from the right with the projector
matrices. Using Wigner's formula for $R^{\Lambda}_{MN}$ (see e.g. [Wi],
chapter XV, formula (27)) one has
$R^{\Lambda}_{MN}({\rm j})=(-1)^{\Lambda+N}\delta_{-M,N}$ and
$R^{\Lambda}_{MN}({\rm  k})=(-1)^N\delta_{MN}$. The projections of the
basis functions can now be written in the final form
$$\eqalignno{
& (T^{\mu}R^{\Lambda})_{MN}
=\shalf(1+r_3(-1)^N)
\left(R^{\Lambda}_{MN}+r_2(-1)^{\Lambda+N}R^{\Lambda}_{M,-N}\right)
&(B.6)\cr
&=\left(R^{\Lambda}_{MN}+r_2(-1)^{\Lambda+N}R^{\Lambda}_{M,-N}\right)
\cases{N\geq 0 \ \hbox{and even} & for $r_3= 1,$\cr
       N\geq 1 \ \hbox{and odd}  & for $r_3=-1.$\cr}
&(B.7)\cr}
$$
These are precisely the bases used in [BM] to describe collective
rotational modes of even-A nuclei. (Compare formula (4-276) in [BM].)
{}From Wigners formula one has
${\b R}^{\Lambda}_{MN}=(-1)^{M-N}R^{\Lambda}_{-M,-N}$, which for
the basis functions
$B^{\Lambda}_{MN}:=R^{\Lambda}_{MN}+r_2(-1)^{\Lambda+N}R^{\Lambda}_{M,-N}$
implies ${\b B}^{\Lambda}_{MN}=r_2(-1)^{\Lambda+M}B^{\Lambda}_{-M,N}$.
This defines the operation of time-reversal -- given by complex
conjugation -- within each sector.

We now turn to the two-dimensional representation $D^4$. Here we
only have to consider $\Lambda={{\rm odd}\over 2}$. Again we
straightforwardly use formula $(4.18b)$ with some normalized
$a\in C^2$ Applied to the functions $R^{\Lambda}_{MN}$, one obtains
$$
T^{4}(a)R^{\Lambda}=R^{\Lambda}P^{\Lambda}(a)\,,
\eqno{(B.8)}
$$
where the right side is again understood as matrix
multiplication with
$$\eqalign{
P^{\Lambda}(a)=\shalf
&  \Big[
   \vert a_1\vert^2 \left({\bf 1}-iR^{\Lambda}({\rm k})\right)
  + \vert a_2\vert^2 \left({\bf 1}+iR^{\Lambda}({\rm k})\right)\cr
&
  + a_1{\b a}_2\left(-iR^{\Lambda}({\rm i})-R^{\Lambda}({\rm j})\right)
+ {\b a}_1a_2\left(-iR^{\Lambda}({\rm i})+R^{\Lambda}({\rm j})\right)
\Big]\,.\cr}
\eqno{(B.9)}
$$
It is not difficult to check explicitly that this is indeed a
projection operator. Using Wigner's
formula\note{We use Wigner's convention which agrees with
$R^{1\over 2}({\rm i,j,k})=-i\tau_{1,2,3}$. It differs from other
conventions in use by a factor $(-1)^{M-N}$, adopted e.g. in [Ha],
formula (9-76).} for $R^{\Lambda}({\rm j})$, $R^{\Lambda}({\rm k})$
and the relation
$R^{\Lambda}({\rm i})=R^{\Lambda}({\rm j})R^{\Lambda}({\rm k})$
we find
$$\eqalignno{
& R^{\Lambda}_{MN}({\rm i}) = \exp(i\pi N)(-1)^{\Lambda+N}\,\delta_{M,-N}
                            = i(-1)^{\Lambda+{1\over 2}}\,\delta_{M,-N}\,,
                              &(B.10)\cr
& R^{\Lambda}_{MN}({\rm j}) = (-1)^{\Lambda+N}\,\delta_{M,-N}\,,
                              &(B.11)\cr
& R^{\Lambda}_{MN}({\rm k}) = \exp(i\pi N)\,\delta_{MN}
                            = (-i)(-1)^{N+{1\over 2}}\,\delta_{MN}\,,
                              &(B.12)\cr}
$$
where the first expressions on the right hand sides are valid for
all $\Lambda$ and the second expressions specialize to
$2\Lambda=\hbox{odd}$ (and hence $2N,2M=\hbox{odd}$). Using them we
obtain
$$\eqalignno{
& \shalf\Big[{\bf 1}\mp iR^{\Lambda}({\rm k})\Big]_{MN}
  =\shalf\left(1\mp (-1)^{N+{1\over 2}}\right)\,\delta_{MN}\,,
& (B.13)\cr
& \shalf\Big[-iR^{\Lambda}({\rm i}) \mp R^{\Lambda}({\rm j})\Big]_{MN}
  = \shalf(-1)^{\Lambda+{1\over 2}}\left(1\mp (-1)^{N-{1\over 2}}\right)\,
  \delta_{M,-N}\,,
& (B.14)\cr}
$$
which inserted into $(B.9)$ gives for $(B.8)$:
$$
T^4(a)R^{\Lambda}_{MN}=
\cases{
 {\b a}_1\left(a_1 R^{\Lambda}_{MN}+
 (-1)^{\Lambda+{1\over 2}} a_2 R^{\Lambda}_{M,-N}\right),
& for $2N= 1\ \hbox{mod}\, 4$\cr
&\cr
 {\b a}_2\left(a_2 R^{\Lambda}_{MN}+
 (-1)^{\Lambda+{1\over 2}} a_1 R^{\Lambda}_{M,-N}\right),
& for $2N=-1\ \hbox{mod}\, 4\,.$\cr}\qquad
\eqno{(B.15a,b)}
$$

If for fixed $\Lambda$ we let $M,N$ run through all $(2\Lambda +1)^2$
values, the right hand side of $(B.15)$ contains $\shalf(2\Lambda+1)^2$
linearly independent functions. For $a_2=0$ they are
$\{R^{\Lambda}_{MN},\ 2N=1\ \hbox{mod}\,4\}$ and
$\{R^{\Lambda}_{MN},\ 2N=-1\ \hbox{mod}\,4\}$ for $a_1=0$.
If $a_1a_2\not =0$ the set of functions in $(B.15a)$ and $(B.15b)$
are the same up to an overall factor. The general representation of
the truncated Hilbert space, ${\H}^{\mu=4}_{\rm tr}$, is therefore
given by
$$\eqalign{
  {\H}^{\mu=4}_{\rm tr}=\hbox{span}
& \Big\{ a_1 R^{\Lambda}_{MN} + (-1)^{\Lambda+N}a_2 R^{\Lambda}_{M,-N}\,,
\cr
&
  2\Lambda=\hbox{odd},\ -\Lambda\leq M,N\leq\Lambda,\
  2N=1\, \hbox{mod}\, 4\Big\}\,,\cr}
\eqno{(B.16)}
$$
where we could set $(-1)^{\Lambda+{1\over 2}}=(-1)^{\Lambda+N}$ due
to $N=\shalf\,\hbox{mod}\,2$. But this is precisely the basis used
in [BM] to describe collective rotational modes for odd-A nuclei.
(Compare formula (4-293) in [BM].)

Finally, if we set $B^{\Lambda}_{MN}(a):=
a_1 R^{\Lambda}_{MN}+(-1)^{\Lambda+N}a_2 R^{\Lambda}_{M,-N}$, we have
${\b B}^{\Lambda}_{MN}(a)=
-(-)^{\Lambda+M}B^{\Lambda}_{-M,N}(i\tau_2\b a)$. Since the canonical
isomorphism $I:{\H}^4_{\rm tr}(a)\rightarrow {\H}^4_{\rm tr}(a')$
is just given by $I(B^{\Lambda}_{MN}(a)):=B^{\Lambda}_{MN}(a')$, the
operation of time reversal, $\T$, can be defined by
${\T}(B^{\Lambda}_{MN}):=(-1)^{\Lambda+M}B^{\Lambda}_{-M,N}$ (plus
antilinear extension) on the set of basis functions $B^{\Lambda}_{MN}$
for fixed $a$.

This concludes the presentation of a relatively simple
example for the usage of non-abelian sectors within familiar quantum
mechanics.

\vskip1.5truecm
\centerline{\bf Acknowledgements}
\smallskip
This work was supported by the Tomalla Foundation Z\"urich
and a research grant from the Center of Geometry and Physics at the
Pennsylvania State University. I thank Petr H\'aj\'{\i}\v cek and Abhay
Ashtekar for their hospitality and discussions.

\vfill\eject

\beginsection{References}

{\parindent=0pt

\item{[Ba1]}
Balachandran, A.P. (1989). Classical Topology and Quantum Phases.
                      In:
                      Geometrical and Algebraic Aspects
                      of Nonlinear Field Theory, S. de Filippo, M. Marinaro,
                      G. Marmo and G. Vilasi (Editors). Elsevier Science
                      Publishers B.V., North Holland.

\item{[Ba2]}
Balachandran, A.P., Marmo, G., Skagerstam, B.S., Stern, A. (1991).
                    Classical Topology and Quantum States. World
                    Scientific Publishing Co., Singapore, New Jersey,
                    London, Hong Kong.

\item{[BLOT]}
Bogolubov, N.N., Logunov, A.A., Oksak, A.I., Todorov, I.T. (1990).
                 General Principles of Quantum Field Theory. Kluwer
                 Academic Publishers, Dordrecht-Boston-London.

\item{[BM]}
Bohr, A., Mottelson, B. (1975). Nuclear Strukture, Vol.II: Nuclear
Deformations. W.A. Benjamin, INC., Reading Massachusetts.


\item{[Di]}
Dirac, P.A.M. (1982). The Principles of Quantum Mechanics.
Fourth edition, Clarendon Press, Oxford.

\item{[Do1]}
Dowker, S. (1972). Quantum Mechanics and Field Theory on Multiply Connected
                   and Homogeneous spaces. {\it J. Phys.} A, {\bf 5}, 936-943.

\item{[Do2]}
Dowker, S. (1979). Selected Topics in Topology and Quantum Field Theory.
                   Lectures delivered at Center for Relativity, Austin,
                   January-May 1979.

\item{[Gi]}
Giulini, D. (1995). On Galilei Invariance in Quantum Mechanics
and the Bargmann Superselection Rule. University of Freiburg, Preprint
Thep 95/15, and quant-ph 9508002.

\item{[GMN]}
Galindo, A., Morales, A., Nu\~nez-Lagos, R. (1962). Superselection
            Principle and Pure States of $n$-Identical Particles.
            {\it Jour. Math. Phys.}, {\bf 3}, 324-328.

\item{[Ha]}
Hamermesh, M. (1964). Group Theory and its Application to Physical
Problems. Addison-Wesley Publ. Comp., Inc. Reading Massachusetts.
Second (corrected) printing.

\item{[HT]}
Hartle, J.B., Taylor, J.R. (1969). Quantum Mechanics and
             Paraparticles. {\it Phys. Rev.}, {\bf 178}, 2043-2051.

\item{[Ho]}
Horuzhy, H. (1976). Superposition Principle in Algebraic Quantum Theory.
                    {\it Theor. Math. Phys.}, {\bf 23}, 413-421.

\item{[Ja]}
Jauch, J.M. (1960). Systems of Observables in Quantum Mechanics.
                    {\it Helv. Phys. Acta} {\bf 33}, 711-726.

\item{[JM]}
Jauch, J.M., Misra, B. (1961). Supersymmetries and Essential
                    Observables. {\it Helv. Phys. Acta} {\bf 34}, 699-709.

\item{[LD]}
Laidlaw, M., DeWitt, C. (1971).
           Feynman Functional Integrals for Systems of Indistinguishable
           Particles. {\ it Phys. Rev.} D, {\bf 3}, 1375-1378.

\item{[MG]}
Messiah, A.M., Greenberg, O.W. (1964). Symmetrization Postulate and
     its Experimental Foundations. {\it Phys. Rev.}, {\bf 136B}, 248-267.

\item{[Re]}
Reeh, H. (1988). A Remark Concerning Canonical Commutation Relations.
                 {\it Jour. Math. Phys.}, {\bf 29}, 1535-1536.

\item{[RS]}
Reed, M., Simon, B. (1972). Methods of Modern Mathematical Physics.
Vol. I: Fuctional Analysis. Academic Press, New York - San Francisco -
London.

\item{[Sch]}
Schr\"odinger, E. (1938). Die Mehrdeutigkeit der Wellenfunktion.
Annalen der Physik (Leipzig) {\bf 32}, 49-55.

\item{[So]}
Sorkin, R. (1989). Classical Topology and Quantum Phases.
                   In:
                   Geometrical and Algebraic Aspects
                   of Nonlinear Field Theory, S. de Filippo, M. Marinaro,
                   G. Marmo and G. Vilasi (Editors). Elsevier Science
                   Publishers B.V., North Holland.



\item{[St]}
Steenrod, N. (1974). The Topology of Fibre Bundles; ninth printing.
Princeton University Press, Princeton, New Jersey.

\item{[T]}
Thirring, W. (1981). A Course in Mathematical Physics, Vol. 3:
Quantum Mechanics of Atoms and Molecules. Springer-Verlag, New York,
Wien.

\item{[We]}
Weyl, H. (1981). Gruppentheorie und Quantenmechanik.
Wissenschaftliche Buchgesellschaft, Darmstadt

\item{[Wi]}
Wigner, E. (1931).
Gruppentheorie und ihre Anwendung auf die Quantenmechanik der
Atomspektren. Friedr. Vieweg \& Sohn Akt.-Ges., Braunschweig.

\item{[Wi1]}
Wightman, A.S. (1959). Relativistic Invariance and Quantum Mechanics
(Notes by A. Barut). {\it Nuovo Cimento, Suppl.}, {\bf 14}, 81-94.

\item{[Wi2]}
Wightman, A.S. (1995). Superselection Rules; Old and New.
{\it Nouvo Cimento}, {\bf 110 B}, 751-769.

\item{[Wo]}
Woodhouse, N. (1980). Geometric Quantization. Claredon Press, Oxford.

}
\vfill\eject
\end